\documentclass[sigconf]{acmart}

\usepackage{multirow}
\usepackage[table,xcdraw]{xcolor}
\usepackage[normalem]{ulem}
\useunder{\uline}{\ul}{}

\usepackage{enumitem}

\usepackage{amssymb}

\usepackage{pifont}
\usepackage{mathtools}
\usepackage{url}
\usepackage{multirow}
\usepackage{subfigure}
\usepackage{bm}
\usepackage{verbatim}
\usepackage{ragged2e}
\usepackage{caption}
\usepackage{subcaption}
\usepackage{graphicx}
\usepackage[normalem]{ulem}
\useunder{\uline}{\ul}{}
\usepackage{amsmath}
\usepackage{longtable}
\usepackage{adjustbox}
\usepackage{microtype}
\usepackage{setspace}
\usepackage{hyperref}
\usepackage{float}
\usepackage{balance}
\usepackage{algorithm}
\usepackage{algpseudocode}
\newcommand{\ie}{\emph{i.e., }}
\newcommand{\eg}{\emph{e.g., }}

\newcommand{\aka}

\AtBeginDocument{%
  \providecommand\BibTeX{{%
    \normalfont B\kern-0.5em{\scshape i\kern-0.25em b}\kern-0.8em\TeX}}}

\copyrightyear{2018}
\setcopyright{acmlicensed}
\acmYear{2018}
\acmDOI{XXXXXXX.XXXXXXX}
\acmConference[Conference acronym 'XX]{Make sure to enter the correct
  conference title from your rights confirmation emai}{June 03--05,
  2018}{Woodstock, NY}
\acmBooktitle{Woodstock '18: ACM Symposium on Neural Gaze Detection,
 June 03--05, 2018, Woodstock, NY} 
\acmISBN{978-1-4503-XXXX-X/18/06}

\begin{document}

\title{UniGRec: Unified Generative Recommendation with Soft Identifiers for End-to-End Optimization}


\author{Jialei Li}
\orcid{0009-0007-9845-0251}
\affiliation{
  \institution{University of Science and Technology of China}
  \city{Hefei}
  \country{China}
}
\email{lijialei.cn@gmail.com}

\author{Yang Zhang}
\orcid{0000-0002-7863-5183}
\affiliation{
  \institution{National University of Singapore}
  \city{Singapore}
  \country{Singapore}
}
\email{zyang1580@gmail.com}

\author{Yimeng Bai}
\orcid{0009-0008-8874-9409}
\affiliation{
  \institution{University of Science and Technology of China}
  \city{Hefei}
  \country{China}
}
\email{baiyimeng@mail.ustc.edu.cn}

\author{Shuai Zhu}
\orcid{}
\affiliation{
  \institution{University of Science and Technology of China}
  \city{Hefei}
  \country{China}
}
\email{zhushuaiozj@mail.ustc.edu.cn}

\author{Ziqi Xue}
\orcid{}
\affiliation{
  \institution{University of Science and Technology of China}
  \city{Hefei}
  \country{China}
}
\email{ziqi@mail.ustc.edu.cn}

\author{Xiaoyan Zhao}
\orcid{0000-0001-6001-1260}
\affiliation{
  \institution{The Chinese University of Hong Kong}
  \city{Hong Kong}
  \country{China}}
\email{xzhao@se.cuhk.edu.hk}

\author{Dingxian Wang}
\orcid{0000-0002-6880-7869}
\affiliation{
  \institution{Upwork}
  \city{Seattle}
  \country{USA}
}
\email{dingxianwang@upwork.com}

\author{Frank Yang}
\orcid{}
\affiliation{
  \institution{Upwork}
  \city{Seattle}
  \country{USA}
}
\email{frankyang@upwork.com}

\author{Andrew Rabinovich}
\orcid{}
\affiliation{
  \institution{Upwork}
  \city{Seattle}
  \country{USA}
}
\email{andrewrabinovich@upwork.com}

\author{Xiangnan He}
\orcid{0000-0001-8472-7992}
\affiliation{
  \institution{University of Science and Technology of China}
  \city{Hefei}
  \country{China}
}
\email{xiangnanhe@gmail.com}



\renewcommand{\shortauthors}{Jialei Li et al.}

\begin{abstract}

Generative recommendation has recently emerged as a transformative paradigm that directly generates target items, surpassing traditional cascaded approaches. It typically involves two components: a tokenizer that learns item identifiers and a recommender trained on them. Existing methods often decouple tokenization from recommendation or rely on asynchronous alternating optimization, limiting full end-to-end alignment. To address this, we unify the tokenizer and recommender under the ultimate recommendation objective via differentiable soft item identifiers, enabling joint end-to-end training. However, this introduces three challenges: training–inference discrepancy due to soft-to-hard mismatch, item identifier collapse from codeword usage imbalance, and collaborative signal deficiency due to an overemphasis on fine-grained token-level semantics.

To tackle these challenges, we propose UniGRec, a unified generative recommendation framework that addresses them from three perspectives. UniGRec employs Annealed Inference Alignment during tokenization to smoothly bridge soft training and hard inference, a Codeword Uniformity Regularization to prevent identifier collapse and encourage codebook diversity, and a Dual Collaborative Distillation mechanism that distills collaborative priors from a lightweight teacher model to jointly guide both the tokenizer and the recommender. Extensive experiments on real-world datasets demonstrate that UniGRec consistently outperforms state-of-the-art baseline methods. Our codes are available at \url{https://github.com/Jialei-03/UniGRec}.

\end{abstract}





\maketitle

\section{Introduction}

Generative recommendation has emerged as a transformative paradigm~\cite{GeneRec,hou2025generative,kuaishou_survey,li-etal-2024-large}, garnering substantial interest across both industry and academia~\cite{TIGER,Onerec-v2,OnePiece,MTGR}. This paradigm typically comprises two components: a tokenizer that maps each item into a sequence of tokens (\ie codewords from predefined codebooks~\cite{TIGER}) serving as its identifier~\cite{RPG,LETTER}, and a recommender that autoregressively predicts the next item based on the tokenized interaction history. By operating within a compact token space, such generative models circumvent the need for exhaustive candidate comparisons, thereby significantly enhancing scalability for massive item corpora~\cite{TIGER} and offering a compelling alternative to the traditional cascaded retrieve-and-rank framework~\cite{StreamingVQ,Trinity,RankMixer}.

Early studies in generative recommendation typically decouples the tokenizer from the recommender, optimizing them in a staged pipeline: the tokenizer is first pre-trained to generate static item identifiers, which then remain frozen during the subsequent training of the recommender~\cite{EAGER,SEATER,TIGER,LETTER,DiscRec,LC-Rec}. However, this unidirectional formulation renders the tokenizer agnostic to downstream recommendation objectives, resulting in rigid identifiers that cannot adapt to the evolving recommendation context~\cite{ETEGRec}. To address this limitation, recent studies have introduced alternating optimization strategies augmented with auxiliary losses, enabling partial coordination between the tokenizer and the recommender~\cite{BLOGER,IDGenRec,ETEGRec}. Nevertheless, due to the absence of a unified high-level objective, this epoch-wise alternation leaves the two components asynchronously updated, preventing their full alignment through genuine end-to-end joint optimization.

\begin{figure}[t]
  \centering
  \includegraphics[width=\linewidth]{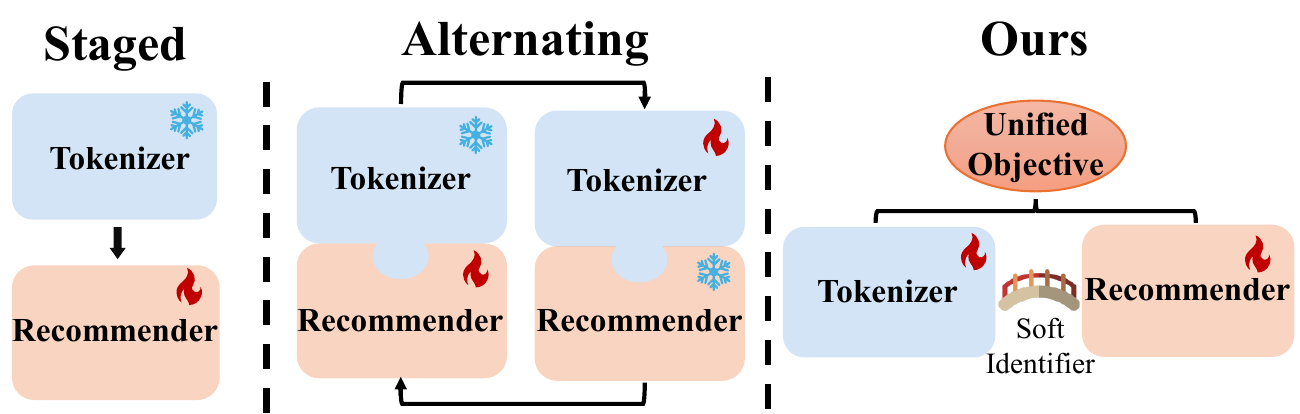}
  \caption{Comparison of generative recommendation training paradigms. Staged training separates optimization into two phases with a frozen tokenizer during recommendation training. Alternating training updates the tokenizer and recommender asynchronously without a unified objective. End-to-end joint training enables a unified gradient flow, jointly optimizing both components under a single objective.}
  \label{fig:paradigm_comparison}
\end{figure}

A natural solution to the limitations of staged and alternating training is to treat the ultimate recommendation loss as a unified objective for the tokenizer, as shown in Figure~\ref{fig:paradigm_comparison}. This can be achieved by substituting discrete codewords (\textit{hard} identifiers) with continuous assignment probabilities (\textit{soft} identifiers)~\cite{BLOGER,MMQ} as the recommender input, which introduces a \textit{differentiable} path that connects the recommendation objective to the tokenizer. On the one hand, this design unifies both components under a single objective, streamlining the training framework and facilitating seamless end-to-end alignment. On the other hand, it substantially mitigates the information loss inherent in hard quantization, enabling the model to preserve and capture richer fine-grained patterns~\cite{IBQ}.

Despite its conceptual appeal, implementing this approach introduces three significant challenges: (1) \textit{Training–Inference Discrepancy}: Continuous soft optimization during training is misaligned with discrete hard inference at test time, which leads to a substantial performance drop during deterministic item generation. (2) \textit{Item Identifier Collapse}: 
With soft identifiers, each item is associated with the entire codebook, amplifying the tendency for optimization to focus on a few dominant codewords, which results in severe codeword usage imbalance and unstable training~\cite{EdVAE,Zhu_2025_ICCV}.
(3) \textit{Collaborative Signal Deficiency}: 
While soft identifiers enhance the capture of fine-grained, token-level semantics, this microscopic focus may come at the expense of modeling coarse-grained, item-level collaborative signals~\cite{BinLLM,CoLLM,DiscRec}, which are indispensable for high-quality recommendation.

To tackle these challenges, we propose an innovative framework named \textbf{UniGRec} (Unified Generative Recommendation). First, to mitigate the training–inference discrepancy, we introduce \textit{Annealed Inference Alignment}, where the codeword assignment logits are temperature-scaled to compute probabilities during tokenization, with the temperature gradually annealed during training. This strategy enables the assignment distribution to smoothly evolve toward a sharp, near-deterministic form, thereby ensuring consistency between training and inference. Second, to address item identifier collapse, we incorporate \textit{Codeword Uniformity Regularization}, which penalizes over-concentration on dominant codewords, encouraging more diverse item identifiers and reducing codebook collisions. Third, to overcome collaborative signal deficiency, we introduce \textit{Dual Collaborative Distillation}, which distills collaborative priors from a lightweight teacher model (\eg SASRec~\cite{SASRec}), by aligning the outputs of both the tokenizer and the recommender with its pre-trained item embeddings. This mechanism explicitly reinforces coarse-grained collaborative signal extraction across items. Overall, UniGRec integrates these components into a differentiable and end-to-end training framework, enabling joint optimization of both models under a unified high-level objective --- recommendation loss. Extensive experiments on multiple real-world datasets demonstrate that UniGRec consistently outperforms existing baseline methods.

The main contributions of this work are summarized as follows:
\begin{itemize}[leftmargin=*]
\item We introduce soft identifiers in generative recommendation, allowing the ultimate recommendation loss to serve as a unified objective, and identify three key challenges arising from this design: Training–Inference Discrepancy, Item Identifier Collapse, and Collaborative Signal Deficiency.
\item We propose UniGRec, which integrates Annealed Inference Alignment, Codeword Uniformity Regularization, and Dual Collaborative Distillation to jointly optimize the tokenizer and recommender in a differentiable, end-to-end training framework.
\item We conduct extensive experiments on multiple real-world datasets, demonstrating the effectiveness of UniGRec.
\end{itemize}
\section{Preliminary}

In this section, we formulate the generative recommendation task from a modeling perspective, focusing on a \emph{sequential recommendation} setting with two key components: a tokenizer and a recommender. Here, we take TIGER~\cite{TIGER} as a representative example to illustrate the framework due to its widespread adoption. Let $\mathcal{I}$ denote the entire set of items. Given a user's interaction sequence $S = [i_1, i_2, \ldots, i_T]$, the objective is to predict the next target item $i_{T+1} \in \mathcal{I}$. Generative recommendation reformulates this problem as directly generating the target item from the tokenized interaction history, rather than retrieving it from a massive candidate set, as in traditional matching-based paradigms~\cite{SASRec}.

\subsection{Item Tokenization}
\subsubsection{Overview}
In this stage, each item is encoded into a sequence of discrete tokens using a tokenizer model $\mathcal{T}$ based on an RQ-VAE~\cite{RQ-VAE}. The tokenizer quantizes continuous item semantic embeddings (\eg usually textual representations~\cite{TIGER,LETTER}) into hierarchical token sequences in a coarse-to-fine manner. Each token corresponds to a codeword from a level-specific codebook, where the first-level codewords capture coarse semantics and deeper-level codewords progressively refine item-specific details.
Formally, given an item $i$ with semantic embedding $\bm{z}$, the tokenizer produces an $L$-level token sequence:
\begin{equation}
[c_1, \ldots, c_L] = \mathcal{T}(\bm{z}),
\end{equation}
where $c_l$ denotes the selected codeword at level $l$.

\subsubsection{Tokenizer}
We next introduce the architecture and optimization of the tokenizer model. First, the tokenizer $\mathcal{T}$ encodes the item semantic embedding $\bm{z}$ into a latent representation using an MLP-based encoder:
\begin{equation}
\bm{r} = \text{Encoder}_{\mathcal{T}}(\bm{z}).
\end{equation}
The latent representation $\bm{r}$ is then discretized into $L$ hierarchical tokens via residual vector quantization with $L$ independent codebooks. For the $l$-th level, the codebook $\bm{\mathcal{C}}_l = \{\bm{e}_l^k\}_{k=1}^{K}$ consists of $K$ codeword embeddings. The token assignment and residual update are defined as:
\begin{align}\label{eq:assgin}
c_l &= \arg\min_k \lVert \bm{v}_l - \bm{e}_l^k \rVert^2, \\
\bm{v}_l &= \bm{v}_{l-1} - \bm{e}_{l-1}^{c_{l-1}},
\end{align}
where $\bm{v}_1 = \bm{r}$ and $\bm{v}_l$ denotes the residual vector at level $l$.
After quantization across all levels, the final discrete representation is obtained by aggregating the selected codeword embeddings:
\begin{equation}
\tilde{\bm{r}} = \sum_{l=1}^{L} \bm{e}_l^{c_l}.
\end{equation}
An MLP-based decoder then reconstructs the semantic embedding:
\begin{equation}
\tilde{\bm{z}} = \text{Decoder}_{\mathcal{T}}(\tilde{\bm{r}}).
\end{equation}

The tokenizer is optimized by minimizing a reconstruction loss together with a quantization loss:
\begin{equation}\label{eq:rqvae_loss}
\mathcal{L}_\text{Token} =
\underbrace{\lVert \tilde{\bm{z}} - \bm{z} \rVert^2}_{\mathcal{L}_\text{Recon}}
+ \underbrace{\sum_{l=1}^{L} \Big(
\lVert \text{sg}(\bm{v}_l) - \bm{e}_l^{c_l} \rVert^2
+ \beta \lVert \bm{v}_l - \text{sg}(\bm{e}_l^{c_l}) \rVert^2
\Big)}_{\mathcal{L}_\text{Quant}}.
\end{equation}
where $\text{sg}(\cdot)$ denotes the stop-gradient operator and $\beta$ is a weighting coefficient (set to $0.25$ by default) that balances the updates between the encoder and the codebooks. The reconstruction term encourages $\tilde{\bm{z}}$ to approximate $\bm{z}$, while the remaining terms enforce consistent quantization of residual representations.

By applying the tokenizer to each item in the interaction history, the input sequence $S$ is transformed into a flattened token sequence:
\begin{equation}
X = [c_1^1, \ldots, c_L^1, \ldots, c_1^{T}, \ldots, c_L^{T}],
\end{equation}
where $c_l^j$ denotes the $l$-th token of item $i_j$. Likewise, the ground-truth next item $i_{T+1}$ is encoded as
\begin{equation}
Y = [c_1^{T+1}, \ldots, c_L^{T+1}].
\end{equation}

\subsection{Autoregressive Generation}

\subsubsection{Overview}
In this stage, next-item prediction is formulated as an autoregressive sequence generation task conditioned on the tokenized interaction history. Given the flattened token sequence $X$, the recommender generates the target item tokens in a left-to-right manner, where each token is predicted based on $X$ and previously generated tokens. This formulation allows the model to capture fine-grained semantic information encoded in item tokens as well as sequential collaborative signals from user interactions.

\subsubsection{Recommender}
We now describe the architecture and optimization of the recommender model $\mathcal{R}$, which adopts a Transformer-based encoder--decoder architecture~\cite{T5}. The input token sequence $X$ is first mapped to embeddings via a shared token embedding table $\bm{\mathcal{O}} \in \mathbb{R}^{|\mathcal{V}|\times D}$:
\begin{equation}\label{eq:rec_input}
\bm{E}^X = [\bm{o}_1^1, \ldots, \bm{o}_L^T],
\end{equation}
where $\bm{o}_l^t$ denotes the embedding of the $l$-th token of item $i_t$, $\mathcal{V}$ is the token vocabulary, and $D$ is the embedding dimension. The encoder processes $\bm{E}^X$ to produce contextualized representations:
\begin{equation}\label{eq:rec_encoder}
\bm{H}^{\text{Enc}} = \text{Encoder}_{\mathcal{R}}(\bm{E}^X).
\end{equation}
For decoding, the decoder generates the target token sequence autoregressively, conditioned on the encoder outputs and the previously generated tokens. During training, a beginning-of-sequence token [BOS] is prepended to the target sequence, and teacher forcing is applied by feeding the ground-truth prefix tokens to the decoder. The decoder produces hidden states as:
\begin{equation}\label{eq:rec_decoder}
\bm{H}^{\text{Dec}} = \text{Decoder}_{\mathcal{R}}(\bm{H}^{\text{Enc}}, \bm{E}^Y),
\end{equation}
where $\bm{E}^Y = [\bm{o}^{\text{BOS}}, \bm{o}_1^{T+1}, \ldots, \bm{o}_L^{T+1}]$.

Each decoder hidden state is projected onto the token vocabulary via inner product with $\bm{\mathcal{O}}$ to predict the next token. The recommender is trained by minimizing the negative log-likelihood of the target token sequence:
\begin{equation}\label{eq:loss_rec}
\mathcal{L}_{\text{Rec}} = -\sum_{l=1}^{L} \log P(Y_l \mid X, Y_{<l}),
\end{equation}
where $Y_l$ denotes the $l$-th token of the target item $i_{T+1}$.

\section{Methodology}
\label{sec:method}

In this section, we first provide an overview of the proposed methodology, followed by a detailed introduction of its components.

\begin{figure*}[t]
  \centering
  \includegraphics[width=0.95\textwidth]{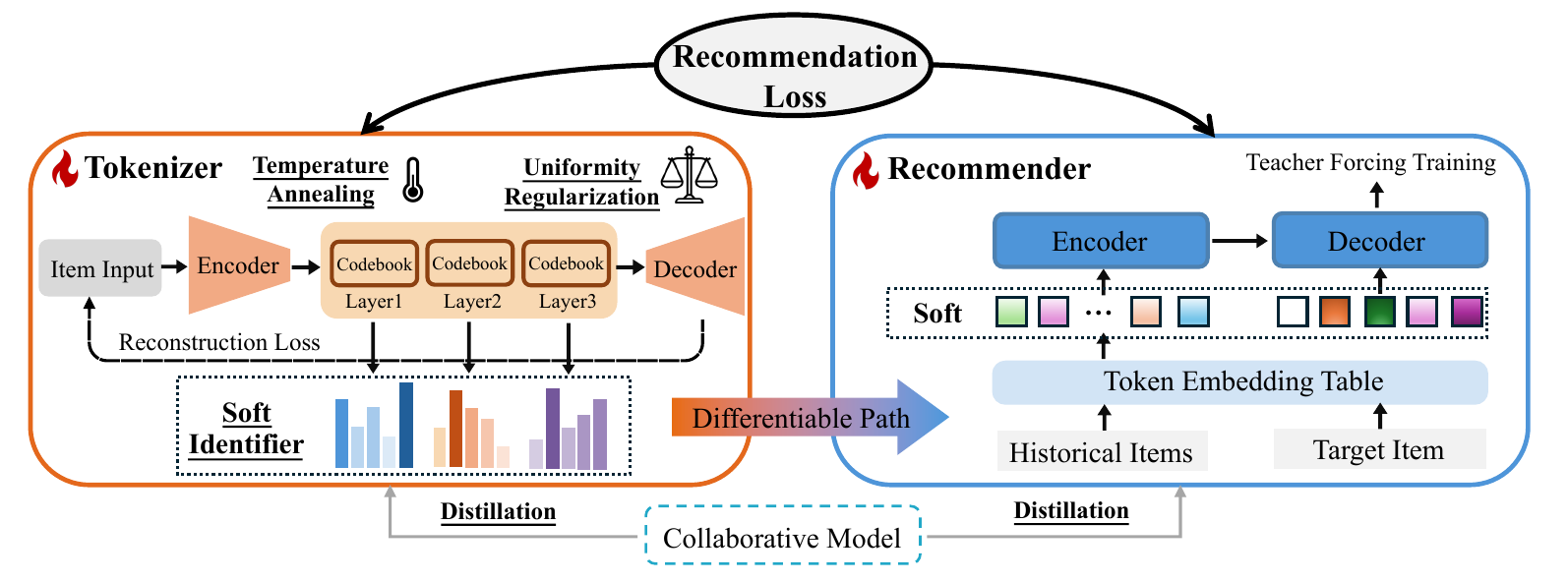}
    \caption{Overview of the proposed UniGRec framework. By introducing soft identifiers, the tokenizer and recommender are seamlessly integrated, enabling end-to-end joint optimization under a unified recommendation objective. Temperature annealing and uniformity regularization are crucial for practical effectiveness, while a pre-trained lightweight teacher model offers additional collaborative supervision.}

  \label{fig:framework}
\end{figure*}

\subsection{Overview}
We aim to unify the tokenizer and the recommender under a single objective and perform end-to-end joint optimization, rather than training them sequentially or optimizing them asynchronously. This design simplifies the training pipeline and leads to better alignment between the two components. Our key belief is that the ultimate recommendation loss can serve as a unified objective for both modules. To overcome the non-differentiability of hard identifiers, we adopt the \textit{soft identifier}, which replace discrete codewords with continuous assignment probabilities.

Based on this idea, we introduce \textbf{UniGRec} (Unified Generative Recommendation), a unified framework for jointly optimizing the tokenizer and the recommender (Figure~\ref{fig:framework}). UniGRec comprises three components: (1) \textit{Annealed Inference Alignment}, which mitigates the training–inference mismatch induced by soft-to-hard identifier transitions; (2) \textit{Codeword Uniformity Regularization}, which alleviates item identifier collapse caused by over-concentration on dominant codewords; and (3) \textit{Dual Collaborative Distillation}, which compensates for insufficient collaborative supervision by transferring collaborative priors from a lightweight teacher model.

\subsection{Annealed Inference Alignment}
We first introduce the formulation of soft identifiers, and then present the annealing strategy applied to the assignment logits to align training and inference process.

\subsubsection{Soft Identifiers}
Our approach builds upon RQ-VAE~\cite{TIGER}. As shown in Equation~\eqref{eq:assgin}, RQ-VAE assigns each item to its nearest codeword at each quantization layer, resulting in a one-hot identifier. While effective, such hard identifiers exhibit two inherent limitations. First, they incur information loss, as an item may be semantically close to multiple codewords but is forced to select only the nearest one. Second, hard assignments impede joint optimization: the $\arg\min$ operation is non-differentiable, preventing gradients from propagating from the recommendation objective back to the tokenizer. Consequently, although the recommender operates on tokenized representations, the two modules cannot be effectively aligned under a unified training objective.

To address these issues, we adopt \textit{soft identifiers}. Instead of selecting a single codeword, we represent each item as a probability distribution over all codewords. Specifically, we treat the negative squared distances between the residual representation $\bm{r}_{l-1}$ and the codewords $\{\bm{e}_l^k\}_{k=1}^{K}$ as logits, apply temperature scaling, and compute assignment probabilities via a softmax function:
\begin{equation}\label{eq:soft_prob}
p_l(k) = \frac{\exp \left( - \| \bm{r}_{l-1} - \bm{e}_l^k \|^2 / \tau \right)}{\sum_{j=1}^{K} \exp \left( - \| \bm{r}_{l-1} - \bm{e}_l^j \|^2 / \tau \right)}, \quad k=1,\cdots, K, 
\end{equation}
where $\tau$ denotes the temperature parameter controlling the sharpness of the assignment distribution. 

Consequently, the discrete code selection is replaced by a probability-weighted aggregation of codewords. These soft assignments preserve richer semantic information and, more importantly, enable fully differentiable end-to-end optimization, allowing the recommendation objective to directly supervise the tokenizer.

To feed soft identifiers into the recommender, we further adapt the input representation. Note that the assignment distribution $p_l(\cdot)$ is defined over $K$ codewords, which is typically much smaller than the full token vocabulary size $|\mathcal{V}|$ of the recommender. Therefore, we first initialize a zero vector with the same dimensionality as the token vocabulary, and then scatter the assignment probabilities corresponding to the codewords of the current quantization layer into their respective token positions. Concretely, we replace the standard embedding lookup with a probability-weighted embedding aggregation. 
Taking the encoder input token $\bm{o}_l^t$ in Equation~\eqref{eq:rec_input} as an example, we compute it as:
\begin{equation}\label{eq:new_rec_input}
\bm{o}_l^t = \text{scatter}\big(\bm{0}, \{p_l(k)\}_{k=1}^{K}\big)\cdot \bm{\mathcal{O}},
\end{equation}
where $\text{scatter}(\cdot)$ maps the $K$-dimensional assignment probabilities to their corresponding indices in the full vocabulary space, producing a sparse $|\mathcal{V}|$-dimensional vector, and $\bm{\mathcal{O}} \in \mathbb{R}^{|\mathcal{V}| \times D}$ denotes the token embedding table of the recommender. This formulation reduces to standard embedding lookup when $p_l(\cdot)$ degenerates to a one-hot vector, while remaining fully differentiable with respect to the tokenizer parameters.

\subsubsection{Temperature Annealing}
Although the proposed design enables end-to-end optimization between the tokenizer and the recommender, it introduces a \textit{Training–Inference Discrepancy}: the recommender is trained with soft identifiers, whereas inference requires deterministic item representations. To mitigate this issue, we apply a linear annealing strategy to the temperature parameter $\tau$ in Equation~\eqref{eq:soft_prob}, progressively reducing it during training:
\begin{equation}\label{eq:tau_sche}
    \tau(\text{step}) = \tau_{\max} - \frac{\text{step}}{\text{total\_step}} \left(\tau_{\max} - \tau_{\min}\right),
\end{equation}
where $\text{step} \in [0, \text{total\_step}]$ denotes the current training step, and $\tau_{\max}$ and $\tau_{\min}$ are the initial (maximum) and final (minimum) temperatures, respectively.

As $\tau$ decreases, the soft assignment distribution becomes increasingly peaked, concentrating probability mass on the nearest codeword. Consequently, the soft weighted embedding gradually converges to the deterministic embedding obtained via hard assignment. This annealed inference alignment enables smooth, fully differentiable optimization in early training, while progressively matching the inference-time behavior.

\subsection{Codeword Uniformity Regularization}

As shown in Equation~\eqref{eq:soft_prob}, each item interacts with the full codebook at the gradient level, which can bias optimization toward a few dominant codewords. In our experiments, this often leads to severe codeword usage imbalance, a phenomenon we term \emph{Item Identifier Collapse}, consistent with prior work~\cite{EdVAE,Zhu_2025_ICCV}. 

To address this, we regularize the codeword assignment probabilities to promote uniform usage, computing batch-level averages to reduce noise and stabilize training. Specifically, we introduce a diversity loss to penalize low-entropy distributions:
\begin{equation}\label{eq:CU}
    \mathcal{L}_\text{CU} = \sum_{l=1}^{L} \sum_{k=1}^{K} \bar{p}_l(k) \log (\bar{p}_l(k)),
\end{equation}
where $\bar{p}_l(k)$ denotes the batch-averaged assignment probability of the $k$-th codeword at the $l$-th quantization level. Minimizing $\mathcal{L}_\text{CU}$ is equivalent to minimizing the KL divergence between $\bar{p}_l$ and a uniform distribution, thereby encouraging more balanced codeword utilization.

\subsection{Dual Collaborative Distillation}

Although soft identifiers are effective at modeling fine-grained, token-level semantics, this emphasis may partially weaken the modeling of coarse-grained, item-level collaborative signals, which provide complementary supervision that is not fully captured by tokenized representations alone~\cite{LETTER,DiscRec}. 

To mitigate this limitation, we introduce a pretrained ID-based collaborative recommender (SASRec~\cite{SASRec}) as a teacher model and leverage its item ID embeddings to supplement collaborative knowledge.
Specifically, we perform collaborative distillation from two complementary perspectives, as detailed below:

\subsubsection{Tokenizer Side}

Our goal on the tokenizer side is to inject collaborative priors into the codebook space, encouraging the soft assignment distributions to align with the quantized representations derived from the collaborative teacher embeddings. To this end, we first extract the historical sequence representation from the recommender encoder, denoted as $\bm{h}^\text{Enc}$ (obtained via average pooling over $\bm{H}^\text{Enc}$ in Equation~\eqref{eq:rec_encoder}), and the unique target item embedding from the pretrained collaborative teacher model, denoted as $\bm{h}^\text{Tea}$. We then project these representations through a linear layer to match the input dimension of the tokenizer and feed them into the tokenizer to obtain the codebook assignment probability distributions, denoted as $P_l^\text{Enc}=\{p_l(k)^\text{Enc}\}_{k=1}^K$ and $P_l^\text{Tea}=\{p_l(k)^\text{Tea}\}_{k=1}^K$, respectively. The subscript $l$ indicates that these distributions correspond to the $l$-th quantization layer.

Finally, we define a symmetric KL divergence loss between the two distributions for each codeword:
\begin{equation}\label{eq:CDT}
    \mathcal{L}_\text{CD}^\mathcal{T} = \sum_{l=1}^{L} \left[ D_{KL}\big(P_l^\text{Enc} \,\|\, P_l^\text{Tea}\big) + D_{KL}\big(P_l^\text{Tea} \,\|\, P_l^\text{Enc}\big) \right],
\end{equation}
where $D_{KL}(\cdot \| \cdot)$ denotes the KL divergence.

\subsubsection{Recommender Side}

On the recommender side, our goal is to inject collaborative priors directly into the item embedding space, encouraging the recommender to produce representations that are consistent with those of a pretrained collaborative teacher. Specifically, for a target item $i_{T+1}$, we extract its representation from the recommender decoder, denoted as $\bm{h}^\text{Dec}_i$ (obtained via first-token pooling from $\bm{H}^\text{Dec}$ in Equation~\eqref{eq:rec_decoder}), and the corresponding embedding from the pretrained teacher model, denoted as $\bm{h}^\text{Tea}_i$. We then project $\bm{h}^\text{Dec}_i$ through a linear layer to match the teacher embedding dimension and employ an in-batch InfoNCE loss to align the recommender outputs with the teacher embeddings.

Formally, for two sets of embeddings, $A = \{\bm{a}_i\}$ and $B = \{\bm{b}_i\}$, we define the InfoNCE loss as
\begin{equation}
\mathcal{L}_\text{InfoNCE}(A, B) 
= - \sum_{i} \log \frac{\exp(\text{sim}(\bm{a}_i, \bm{b}_i) / \tau')}
{\sum_{j} \exp(\text{sim}(\bm{a}_i, \bm{b}_j) / \tau')},
\end{equation}
where $\text{sim}(\cdot, \cdot)$ denotes the cosine similarity function and $\tau'$ is a temperature hyperparameter distinct from that used elsewhere.
To enforce bidirectional alignment, we define the recommender-side collaborative distillation loss as:
\begin{equation}\label{eq:CDR}
\mathcal{L}_\text{CD}^\mathcal{R} =
\mathcal{L}_\text{InfoNCE}(\{\bm{h}_i^\text{Dec}\}, \{\bm{h}_i^\text{Tea}\})
+ \mathcal{L}_\text{InfoNCE}(\{\bm{h}_i^\text{Tea}\}, \{\bm{h}_i^\text{Dec}\}),
\end{equation}
where the first term treats the recommender embeddings as queries and the teacher embeddings as keys, and the second term reverses the roles to achieve symmetric distillation.

\begin{algorithm}[t]
\caption{Training Pipeline of UniGRec}
\label{alg:training}
\begin{algorithmic}[1]

\State \textbf{Initialize:} Tokenizer $\mathcal{T}$, Recommender $\mathcal{R}$;

\State \textbf{// Stage 1: Tokenizer Pretraining}
\While{not converged}
    \State Sample a mini-batch of items;
    \State Update temperature $\tau$ via Equation~\eqref{eq:tau_sche};
    \State Compute assignment probabilities $p_l(k)$ via Equation~\eqref{eq:soft_prob};
    \State Compute reconstruction loss $\mathcal{L}_\text{Recon}$ 
    via Equation~\eqref{eq:rqvae_loss};
    \State Compute regularization term $\mathcal{L}_\text{CU}$ via Equation~\eqref{eq:CU};
    \State Update $\mathcal{T}$ by minimizing $\mathcal{L}_\text{Pre}$ in Equation~\eqref{eq:pre};
\EndWhile

\State \textbf{// Stage 2: End-to-End Joint Training}
\State Fix temperature $\tau \gets \tau_{\min}$;
\While{not converged}
    \State Sample a mini-batch of interactions;

    \State Compute reconstruction loss $\mathcal{L}_\text{Recon}$ 
    via Equation~\eqref{eq:rqvae_loss};

    \State Compute inputs of soft identifier via Equation~\eqref{eq:new_rec_input};

    \State Compute recommendation loss $\mathcal{L}_\text{Rec}$ via Equation~\eqref{eq:loss_rec};

    \State Compute tokenizer-side distillation $\mathcal{L}_\text{CD}^\mathcal{T}$ via Equation~\eqref{eq:CDT};
    
    \State Compute recommender-side distillation $\mathcal{L}_\text{CD}^\mathcal{R}$ Equation~~\eqref{eq:CDR};

    \State Update $\mathcal{T}, \mathcal{R}$ by minimizing $\mathcal{L}_\text{Joint}$ in Equation~\eqref{eq:joint};
\EndWhile

\end{algorithmic}
\end{algorithm}

\subsection{Training Pipeline}\label{sec:pipeline}

We summarize the overall training procedure of UniGRec in Algorithm~\ref{alg:training}, which proceeds in two progressive stages:

\subsubsection{Stage 1: Tokenizer Pretraining} 
In practice, the tokenizer generally requires substantially more epochs to converge than the recommender~\cite{TIGER,LETTER}. To address this, we first perform a pretraining phase to stabilize the tokenizer. During this stage, only the tokenizer is optimized, with the objective of establishing a semantically rich and balanced codebook. Omitting this phase can lead to severe codeword collisions, hindering effective joint training. The training objective is defined as:
\begin{equation}\label{eq:pre}
    \mathcal{L}_\text{Pre} = \mathcal{L}_\text{Recon} + \lambda_\text{CU} \, \mathcal{L}_\text{CU},
\end{equation}
where $\mathcal{L}_\text{Recon}$ (Equation~\eqref{eq:rqvae_loss}) ensures reconstruction fidelity, and $\mathcal{L}_\text{CU}$ (Equation~\eqref{eq:CU}) encourages uniform codeword utilization.

Compared to the standard RQ-VAE, our pretraining loss does not include the quantization term, \ie, $\mathcal{L}_\text{Quant}$ (Equation~\eqref{eq:rqvae_loss}). This is because the use of soft identifiers allows the reconstruction loss alone to provide fully differentiable supervision for updating the codebook, eliminating the need for an additional quantization loss.

\subsubsection{Stage 2: End-to-End Joint Training} 
After stabilizing the tokenizer, we perform end-to-end joint optimization of both models. The total training objective is formulated as:
\begin{equation}\label{eq:joint}
    \mathcal{L}_\text{Joint} = \mathcal{L}_\text{Rec} 
    + \lambda_\text{Recon}\, \mathcal{L}_\text{Recon} 
    + \lambda_\text{CD}^\mathcal{T} \, \mathcal{L}_\text{CD}^\mathcal{T} 
    + \lambda_\text{CD}^\mathcal{R} \, \mathcal{L}_\text{CD}^\mathcal{R},
\end{equation}
where $\mathcal{L}_\text{Rec}$ is the recommendation loss defined in Equation~\eqref{eq:loss_rec}, and $\mathcal{L}_\text{CD}^\mathcal{T}$ and $\mathcal{L}_\text{CD}^\mathcal{R}$ denote the collaborative distillation losses for the tokenizer and recommender, respectively. 

Note that the codeword uniformity regularization $\mathcal{L}_\text{CU}$ is omitted in this stage. This is because, after pretraining, the assignment distributions have already become sufficiently sharp, and further regularization is no longer necessary. We fix the temperature at its minimum value $\tau_{\min}$ to stabilize training while ensuring that the tokenizer and recommender embeddings remain well aligned with the recommendation objective.

\section{Experiment}
\label{sec:experiments}

In this section, we conduct a series of experiments to answer the following research questions:

\noindent \textbf{RQ1.} How does UniGRec perform in comparison with existing generative recommendation methods?

\noindent \textbf{RQ2.} What is the contribution of each individual component to the overall effectiveness of UniGRec?

\noindent \textbf{RQ3.} How do specific hyperparameters or designs affect the performance of UniGRec?

\noindent \textbf{RQ4.} What are the underlying factors driving the observed performance gains of UniGRec?

\subsection{Experimental Setting}

\begin{table}[t]
\caption{Statistical details of the evaluation datasets, where “AvgLen” is the average length of historical sequences. }
\label{table:datasets}
\begin{tabular}{cccccc}
\hline
Dataset       & \#User & \#Item & \#Interaction & Sparsity & AvgLen \\ \hline
Beauty        & 22362  & 12083  & 198313        & 99.93\%  & 8.87 \\
Pet           & 19855  & 8498   & 157747        & 99.91\%  & 7.95 \\
Upwork        & 15542  & 33508  & 139217        & 99.97\%  & 8.96 \\ \hline
\end{tabular}
\end{table}

\subsubsection{Datasets}

We conduct experiments on three real-world datasets, including two publicly available benchmarks from the Amazon Reviews corpus\footnote{\url{https://jmcauley.ucsd.edu/data/amazon/index_2014.html}} (\textit{Beauty} and \textit{Pet}), and one private production dataset collected from our online freelancing platform (\textit{Upwork}). In the Upwork dataset, each interaction corresponds to a time-stamped employer–freelancer hiring event, where employers are treated as users and freelancers are treated as items. The detailed statistics of all datasets are summarized in Table~\ref{table:datasets}.

Following prior work~\cite{SASRec, TIGER}, we apply a 5-core filtering strategy, retaining only users and items with at least five interactions. Each user’s interaction history is truncated or padded to a fixed length of 20 using the most recent interactions. We adopt a leave-one-out splitting strategy, where the most recent interaction of each user is used for testing, the second most recent for validation, and the remaining interactions for training.

\subsubsection{Baselines}
\label{sec:exp_baselines}

The baseline methods used for comparison fall into the following two categories:

\vspace{+3pt}
(a) \noindent\textit{Traditional recommendation methods:}
\begin{itemize}[leftmargin=*]
    \item \textbf{Caser}~\cite{Caser}: This method leverages both horizontal and vertical convolutional filters to extract sequential patterns from user behavior data, capturing diverse local features.
    \item \textbf{GRU4Rec}~\cite{GRU4Rec}: This method employs recurrent neural network (GRU) for session-based recommendation.
    \item \textbf{SASRec}~\cite{SASRec}: This method leverages the self-attention mechanism in Transformers to capture users’ historical preferences.
    \item \textbf{BERT4Rec}~\cite{BERT4Rec}: This method employs bidirectional self-attention to model user behavior sequences and uses a Cloze objective to predict masked items. 
    \item \textbf{HGN}~\cite{HGN}: This method introduces a hierarchical gating mechanism to adaptively integrate long-term and short-term user preferences derived from historical item sequences.
\end{itemize}

(b) \noindent\textit{Generative recommendation methods:}
\begin{itemize}[leftmargin=*]

    \item \textbf{TIGER}~\cite{TIGER}: This method employs an RQ-VAE to encode item representations into discrete semantic IDs and adopts a generative retrieval paradigm for sequential recommendation.
    
    \item \textbf{LETTER}~\cite{LETTER}: This method extends TIGER by introducing a learnable tokenizer that incorporates hierarchical semantics and collaborative signals, while promoting diversity in code assignment.
    
    \item \textbf{EAGER}~\cite{EAGER}: This method applies a two-stream generative recommender that decouples heterogeneous information into separate decoding paths for parallel modeling of semantic and collaborative signals.

    \item \textbf{OneRec}~\cite{zhou2025openonerec}: This method leverages RQ-Kmeans to learn hierarchical indexing for improved efficiency.
    
    \item \textbf{ETEG-Rec}~\cite{ETEGRec}: This method introduces two auxiliary losses to align the intermediate representations of the tokenizer and recommender, and adopts an alternating optimization strategy to jointly train both components.
    
    \item \textbf{DiscRec-T}~\cite{DiscRec}: This method introduces item-level position embeddings and a dual-branch module to disentangle collaborative and semantic signals at the embedding layer.
\end{itemize}

\subsubsection{Evaluation Metrics}
\label{sec:exp_metrics}
We adopt two widely used metrics for top-$K$ recommendation: Recall@$K$ and NDCG@$K$ with $K \in \{5, 10\}$. To ensure an unbiased evaluation, we perform {full ranking} over the entire item set instead of sampling negatives. For all generative methods, we use a {constrained beam search}~\cite{LETTER} with a beam size of 30 during inference.

\subsubsection{Implementation Details}
\label{sec:exp_details}

For traditional baselines, we use the implementations provided in the open-source library RecBole~\cite{recbole}. For generative baselines, we strictly follow the model configurations and hyperparameter settings specified in their original papers to ensure a fair comparison. For our proposed \textit{UniGRec}, we adopt T5 as the recommender. The initial item embeddings (384 dimensions) are extracted using {sentence-t5-base} (768 dimensions) based on item titles and descriptions for the Beauty and Pet datasets, while for the Upwork dataset, we directly use item embeddings generated by our deployed online model. The tokenizer employs an RQ-VAE with encoder dimensions $\{512, 256, 128, 64\}$ and a codebook comprising three layers, each containing 256 codewords of dimension 32. To ensure unique item identification, we append a dedicated token to each item, following prior work~\cite{TIGER,ETEGRec}.

During tokenizer pre-training, we use a batch size of 1024 and a learning rate of $1\text{e-}3$, with the temperature $\tau$ in Equation~\eqref{eq:soft_prob} annealed from $\tau_{\max} \in \{0.01, 0.05\}$ to $\tau_{\min} = 0.001$. The loss weight $\lambda_\text{CU}$ for codeword uniformity regularization in Equation~\eqref{eq:pre} is set to $1\text{e-}4$. During end-to-end joint training, we employ the AdamW~\cite{AdamW} optimizer with a weight decay of 0.05 and a batch size of 512. The learning rate is tuned within $\{2\text{e-}3, 5\text{e-}3, 8\text{e-}3\}$ for the T5 backbone and $\{2\text{e-}7, 2\text{e-}8\}$ for the differentiable tokenizer in Equation~\eqref{eq:joint}. The reconstruction loss weight $\lambda_\text{Recon}$ is set to 0.5, and the alignment coefficients $\lambda_\text{CD}^\mathcal{T}$ and $\lambda_\text{CD}^\mathcal{R}$ are searched within $\{0.01, 0.05, 0.1, 0.5, 1.0\}$. 
The temperature $\tau^\prime$ in Equation~\eqref{eq:CDR} is set to 0.07.

\begin{table*}[t]
\caption{The overall performance comparisons between the baselines and UniGRec. The best results are highlighted in bold, while the second-best are underlined.}
\label{table:main_result}
\centering 
\resizebox{\textwidth}{!}{%
\begin{tabular}{cccccccccccccc}
\hline
                          &                          & \multicolumn{5}{c}{Traditional Method}                                   & \multicolumn{7}{c}{Generative Method}                                                                    \\ \cmidrule(lr){3-7} \cmidrule(lr){8-14}
\multirow{-2}{*}{Dataset} & \multirow{-2}{*}{Metric} & Caser  & GRU4Rec & SASRec       & BERT4Rec & HGN    & TIGER  & LETTER & EAGER  & OneRec & ETEG-Rec      & DiscRec-T & UniGRec                        \\ \hline
                          & Recall@5                 & 0.0275 & 0.0328  & 0.0531       & 0.0191   & 0.0325 & 0.0432 & 0.0405 & 0.0333 & 0.0398 & {\ul 0.0554} & 0.0501    & {\color[HTML]{000000} \textbf{0.0587}} \\
                          & Recall@10                & 0.0455 & 0.0537  & 0.0872       & 0.0321   & 0.0516 & 0.0701 & 0.0730 & 0.0508 & 0.0649 & {\ul 0.0913} & 0.0853    & {\color[HTML]{000000} \textbf{0.0967}} \\
                          & NDCG@5                   & 0.0173 & 0.0204  & 0.0322       & 0.0119   & 0.0208 & 0.0278 & 0.0252 & 0.0220 & 0.0252 & {\ul 0.0346} & 0.0318    & {\color[HTML]{000000} \textbf{0.0377}} \\
\multirow{-4}{*}{Upwork}  & NDCG@10                  & 0.0231 & 0.0272  & 0.0432       & 0.0161   & 0.0270 & 0.0364 & 0.0356 & 0.0277 & 0.0332 & {\ul 0.0461} & 0.0431    & {\color[HTML]{000000} \textbf{0.0499}} \\ \hline
                          & Recall@5                 & 0.0252 & 0.0397  & 0.0284       & 0.0322   & 0.0331 & 0.0432 & 0.0444 & 0.0399 & 0.0375 & {\ul 0.0525} & 0.0513    & {\color[HTML]{000000} \textbf{0.0548}} \\
                          & Recall@10                & 0.0415 & 0.0601  & 0.0550       & 0.0478   & 0.0569 & 0.0617 & 0.0699 & 0.0536 & 0.0567 & {\ul 0.0809} & 0.0765    & {\color[HTML]{000000} \textbf{0.0825}} \\
                          & NDCG@5                   & 0.0155 & 0.0270  & 0.0172       & 0.0219   & 0.0200 & 0.0281 & 0.0288 & 0.0289 & 0.0251 & {\ul 0.0357} & 0.0346    & {\color[HTML]{000000} \textbf{0.0368}} \\
\multirow{-4}{*}{Beauty}  & NDCG@10                  & 0.0208 & 0.0336  & 0.0244       & 0.0270   & 0.0277 & 0.0346 & 0.0370 & 0.0333 & 0.0312 & {\ul 0.0448} & 0.0427    & {\color[HTML]{000000} \textbf{0.0457}} \\ \hline
                          & Recall@5                 & 0.0210 & 0.0300  & 0.0237       & 0.0199   & 0.0304 & 0.0355 & 0.0357 & 0.0282 & 0.0334 & {\ul 0.0450} & 0.0390    & {\color[HTML]{000000} \textbf{0.0470}} \\
                          & Recall@10                & 0.0369 & 0.0504  & 0.0425       & 0.0337   & 0.0519 & 0.0563 & 0.0590 & 0.0404 & 0.0539 & {\ul 0.0693} & 0.0624    & {\color[HTML]{000000} \textbf{0.0701}} \\
                          & NDCG@5                   & 0.0132 & 0.0188  & 0.0142       & 0.0124   & 0.0186 & 0.0239 & 0.0231 & 0.0191 & 0.0216 & {\ul 0.0292} & 0.0257    & {\color[HTML]{000000} \textbf{0.0315}} \\
\multirow{-4}{*}{Pet}     & NDCG@10                  & 0.0183 & 0.0254  & 0.0202       & 0.0169   & 0.0255 & 0.0305 & 0.0306 & 0.0230 & 0.0282 & {\ul 0.0370} & 0.0333    & {\color[HTML]{000000} \textbf{0.0389}} \\ \hline
\end{tabular}%
}
\end{table*}

\subsection{Overall Performance Comparison (RQ1)}
The overall performance comparison is reported in Table~\ref{table:main_result}. Based on the experimental results, we have the following observations:

\begin{itemize}[leftmargin=*]
    \item \textbf{UniGRec demonstrates superior performance over all baselines on all evaluated datasets.} This superiority can be attributed to the unified end-to-end optimization enabled by \textit{soft identifiers}, which make the tokenizer directly supervised by the recommendation objective, thereby validating the effectiveness of unified optimization for generative recommendation.
    \item \textbf{ETEG-Rec achieves the second-best performance.} By adapting the tokenizer during recommendation, it generates identifiers that are more aligned with the recommendation objective than staged methods. Nevertheless, its reliance on asynchronous updates prevents full alignment between two models, whereas UniGRec leverages fully differentiable and synchronous optimization to achieve superior integration.
    \item \textbf{Generative methods generally outperform traditional ID-based baselines.} This suggests that generative approaches can capture collaborative signals while leveraging multimodal information, thereby validating the effectiveness of the generative paradigm in recommendation.
\end{itemize}

\subsection{Ablation Study (RQ2)}
\label{sec:ablation}

\begin{table}[t]
\centering
\caption{Ablation study of UniGRec components on Beauty.}
\label{table:ablation}
\renewcommand{\arraystretch}{1.1}
\setlength{\tabcolsep}{2.5pt}
\resizebox{\columnwidth}{!}{%
    \begin{tabular}{lcccc}
    \hline
    {Method}  & {Recall@5} & {Recall@10} & {NDCG@5} & {NDCG@10}\\ \hline
    M0: TIGER  & 0.0432 & 0.0617 & 0.0281 & 0.0346\\
    M1: M0 + Soft Identifier  & 0.0464 & 0.0697 & 0.0310 & 0.0385\\
    M2: M1 + Joint Training  & 0.0489 & 0.0742 & 0.0333 & 0.0410\\
    M3: M2 + $\mathcal{L}_\text{CU}$  & 0.0503 & 0.0746 & 0.0337 & 0.0415\\
    M4: M3 + $\mathcal{L}_\text{CD}^\mathcal{T}$  & 0.0509 & 0.0760 & 0.0345 & 0.0425\\
    M5: M3 + $\mathcal{L}_\text{CD}^\mathcal{R}$  & 0.0540 & 0.0810 & 0.0366 & 0.0453\\
    \textbf{M6: UniGRec (Full)}  & \textbf{0.0548} & \textbf{0.0825} & \textbf{0.0368} & \textbf{0.0457} \\ \hline
    \end{tabular}%
}
\end{table}

Our proposed {UniGRec} framework establishes a fully differentiable path for end-to-end joint optimization. Its core components include: (1) \textit{Soft Identifiers} to achieve joint training and bridge the training–inference gap, (2) \textit{Codeword Uniformity Regularization} to prevent codebook collapse, and (3) \textit{Dual Collaborative Distillation} to incorporate additional collaborative signals. To evaluate the contribution of each component, we perform an ablation study on the Beauty dataset, progressively adding each module to the simplest TIGER baseline.
The results are summarized in Table~\ref{table:ablation}, where \textbf{M0} denotes the TIGER baseline, and the subsequent variants (\textbf{M1–M6}) correspond to models with progressively added modules. We analyze the contribution of each component below:
\begin{itemize}[leftmargin=*]
    \item \textbf{Impact of Soft Identifiers and Joint Training (M0 $\rightarrow$ M1 $\rightarrow$ M2).} Adding soft identifiers improves performance by using the full probability distribution, which reduces information loss from discrete quantization. Joint training (M2) further boosts performance, as the tokenizer and recommender are optimized together under a single recommendation loss, allowing the two components to work synergistically.
    \item \textbf{Impact of Codeword Uniformity Regularization (M2 $\rightarrow$ M3).} Adding codeword uniformity regularization improves performance by avoiding over-concentration on dominant codewords. By maximizing the entropy of batch-averaged assignments during pretraining, it effectively alleviates item identifier collapse.
    \item \textbf{Impact of Dual Collaborative Distillation (M3 $\rightarrow$ M6)}. Adding dual collaborative distillation improves performance over M3, with both components together yielding the best results. This validates the effectiveness of using pure collaborative teacher models to distill additional collaborative information.
\end{itemize}

\subsection{Hyperparameter and Design Analysis (RQ3)}
\label{sec:hyper_param}

We next investigate the impact specific designs and hyperparameters in our proposed UniGRec on Beauty. 

\begin{figure}[t]
    \centering
    \includegraphics[width=0.9\columnwidth]{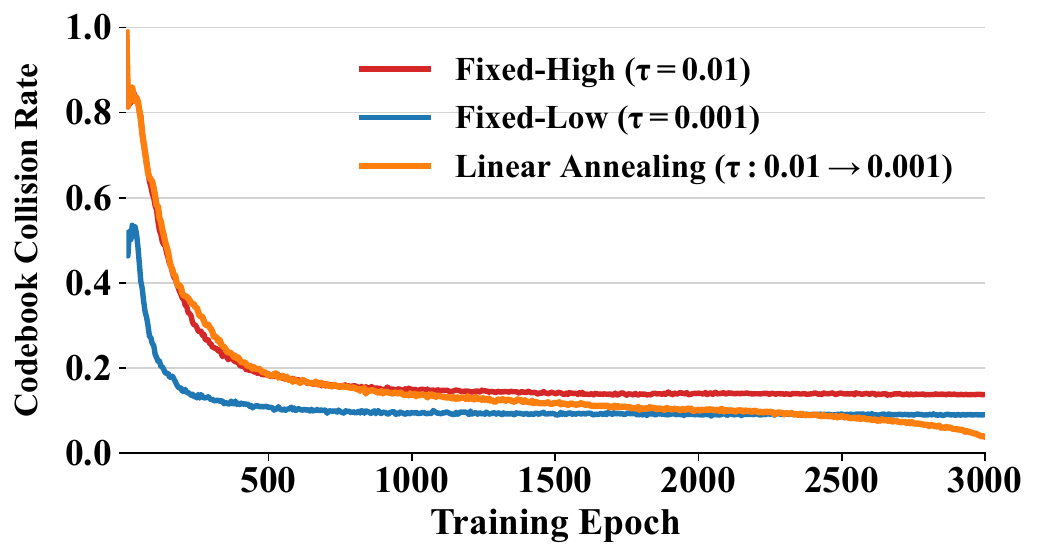}
    \caption{Results of the collision rate curves during training under different temperature ($\tau$) schedules.}
    \label{fig:annealing}
\end{figure}

\subsubsection{Impact of Temperature Annealing Design}
To address the \textit{Training–Inference Discrepancy}, we propose an \textit{Annealed Inference Alignment} strategy that linearly decays the temperature $\tau$ from 0.01 to 0.001. We compare this strategy against two fixed-temperature baselines: Fixed-High ($\tau=0.01$) and Fixed-Low ($\tau=0.001$). We then analyze the codebook collision rate, which is defined as one minus the ratio of the number of unique hard identifiers to the total number of items, over the course of training for each strategy.

The results are summarized in Figure~\ref{fig:annealing}. Fixed-Low maintains a lower collision rate than Fixed-High due to harder assignments, while Fixed-High remains consistently high. Our annealed strategy starts similar to Fixed-High, allowing gradient flow and semantic exploration, then gradually decreases and ultimately falls below Fixed-Low. This shows that annealed inference alignment enables both early exploration and minimal codebook collisions, effectively bridging the training–inference gap.

\begin{figure}[t]
    \centering
    \includegraphics[width=\linewidth]{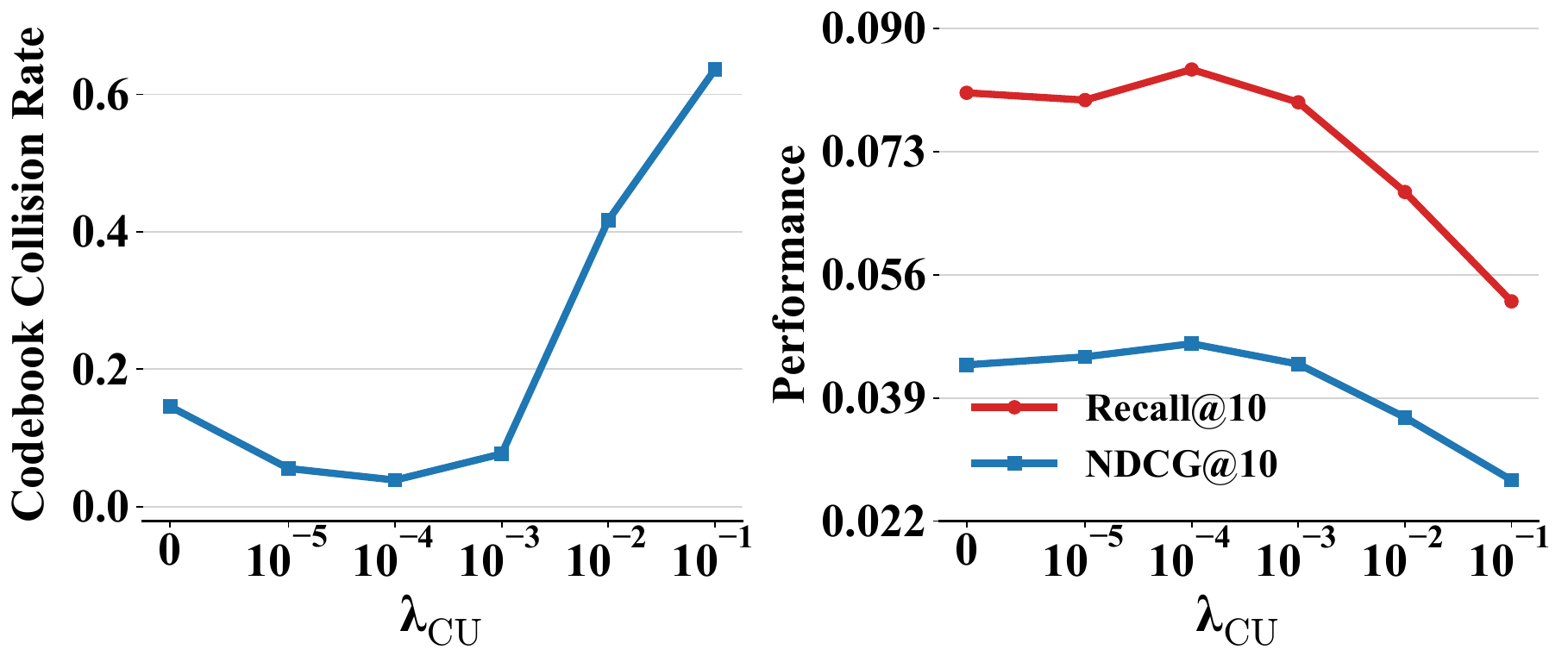}
    \caption{Results of the collision rate and recommendation performance of UniGRec across different values of the uniformity regularization weight ($\lambda_\text{CU}$). }
    \label{fig:diversity}
\end{figure}

\subsubsection{Impact of Hyperparameter $\lambda_{\text{CU}}$}

To mitigate \textit{Item Identifier Collapse}, we introduce \textit{Codeword Uniformity Regularization} ($\mathcal{L}_\text{CU}$) and vary its weight $\lambda_{\text{CU}}$ over $[0, 10^{-5}, \dots, 10^{-1}]$, measuring the resulting codebook collision rate and recommendation performance.

Figure~\ref{fig:diversity} shows a strong negative correlation between codebook collision rate and recommendation accuracy. As $\lambda_\text{CU}$ increases from $0$ to $10^{-4}$, collisions are effectively reduced, resulting in peak performance and confirming that a more balanced codebook enhances item differentiation. However, when $\lambda_\text{CU}$ exceeds $10^{-2}$, excessive regularization causes collisions to rise sharply and performance to decline, likely due to disruption of the semantic embedding reconstruction process.

\subsection{In-Depth Analysis (RQ4)}
\label{sec:analysis}

\begin{figure}[t]
  \centering
  \includegraphics[width=\linewidth]{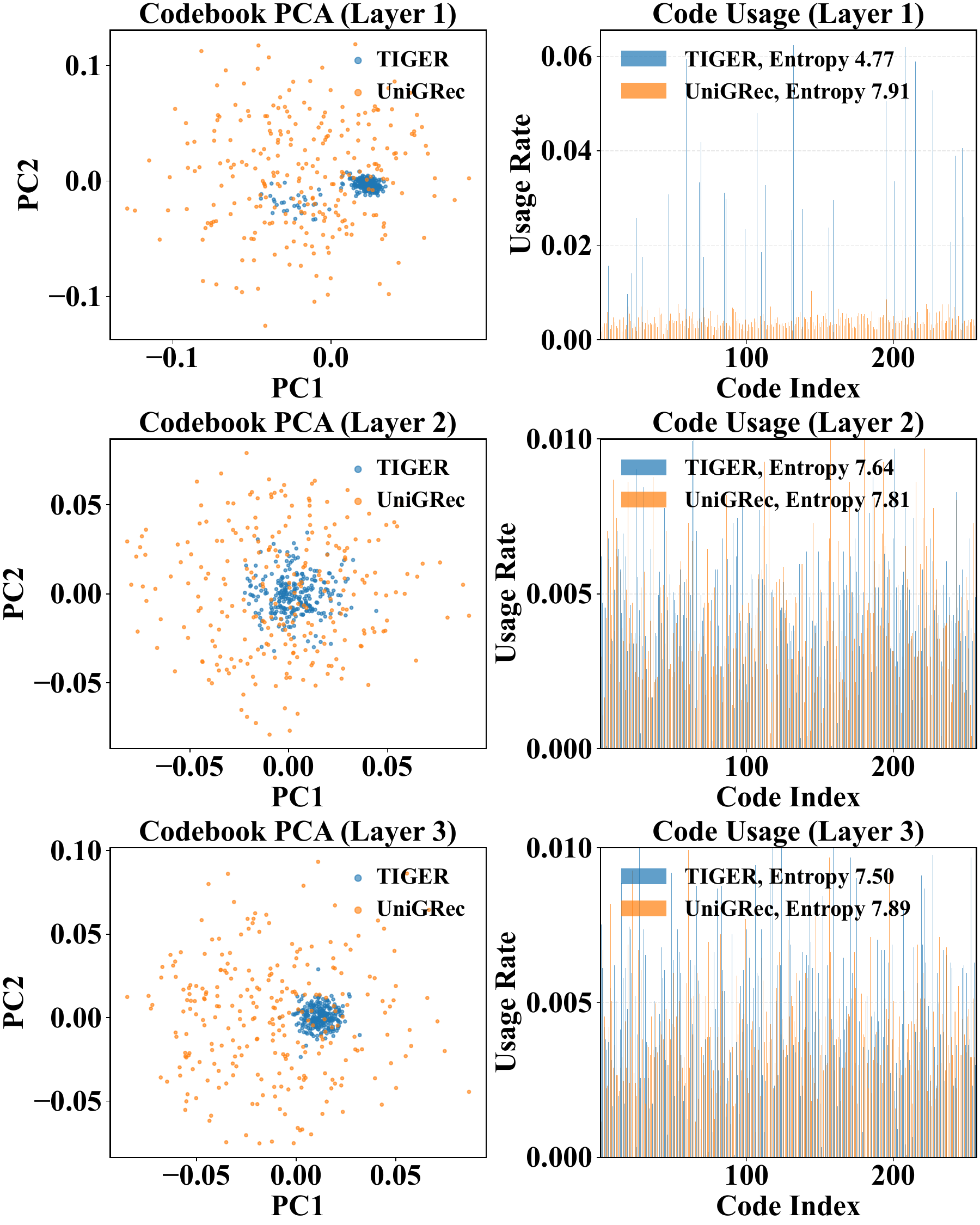}
  \caption{Visualization analysis of codebook embeddings and codeword usage of each quantization layer, where “Entropy” measures the distribution of codeword usage, indicating how evenly the codewords are utilized.}
  \label{fig:codebook_analysis}
\end{figure}

Finally, we conduct a detailed analysis to understand the sources of improvement and identify the specific changes responsible for the observed gains.

\subsubsection{Codebook Analysis}
We examine the codeword embeddings across the three layers of the codebook. For each layer, we apply PCA to reduce the embeddings to two dimensions for visualization. In addition, we analyze codeword usage across all items based on their hard identifiers, computing the entropy to quantify how evenly the codewords are utilized. For comparison, we include results for TIGER, which serves as an ablated version of our framework with all specialized design components removed.

The results, summarized in Figure~\ref{fig:codebook_analysis}, reveal clear differences between the methods. From the perspective of representation, UniGRec exhibits a widely dispersed, isotropic distribution that effectively spans the latent space, indicating a more expressive semantic space where items are mapped to distinct identifiers that are easier for the recommender to differentiate. From the perspective of utilization, UniGRec achieves a near-uniform codeword distribution with substantially higher entropy, whereas TIGER concentrates usage on a few dominant codewords. These findings demonstrate that our method effectively mitigates codebook collapse and maximizes representational capacity.

\subsubsection{Identifier Analysis}

After analyzing the differences between UniGRec and TIGER at the codebook level, we further investigate the training dynamics of UniGRec by examining how item identifiers evolve across the two training stages (\ie, stage~1 and stage~2 described in Section~\ref{sec:pipeline}). For clarity, we focus on hard identifiers, which are derived from the corresponding soft identifiers. We extract the identifiers of all items after stage~1 and stage~2 training, respectively, and conduct a statistical analysis from two perspectives. First, we measure the proportion of identifier changes at each layer. Second, from an item-level perspective, we analyze the change patterns, \ie, which layers of an item’s identifier are modified.

\begin{figure}[t]
  \centering
  \includegraphics[width=\linewidth]{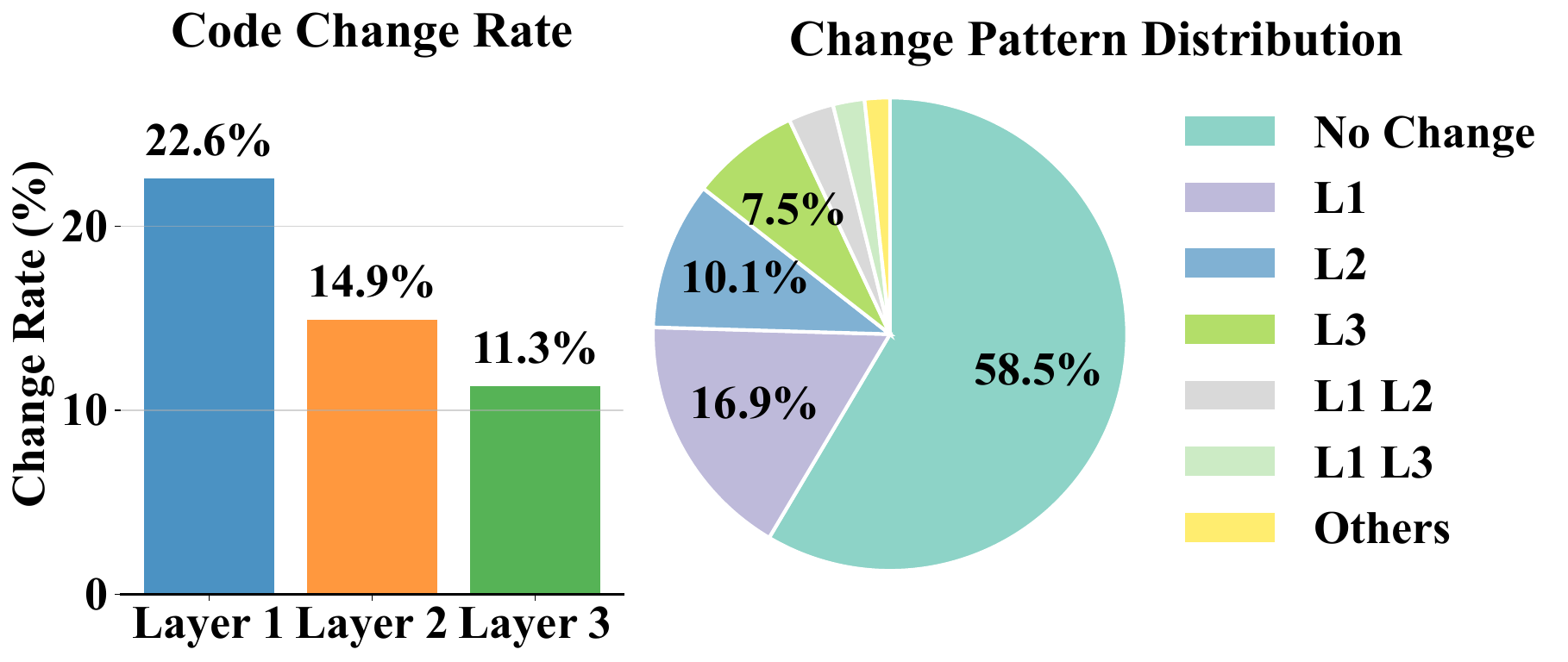}
  \caption{Analysis of identifier evolution, including the code change rate across layers and the distribution of identifier change patterns.}
  \label{fig:case_study}
\end{figure}

The results are summarized in Figure~\ref{fig:case_study}, from which we draw the following observations:
\begin{itemize}[leftmargin=*]
\item \textbf{Layer-level analysis.} The change rate remains relatively stable and low overall, and gradually decreases with increasing layer depth, which aligns with the design principle of residual quantization. This suggests that joint training refines item identifiers in a coarse-to-fine manner.
\item \textbf{Item-level analysis.} Over 93\% of items exhibit changes in at most one layer or remain unchanged. This indicates that end-to-end joint training acts as a recommendation-aware fine-tuning of the tokenizer rather than a disruptive retraining. Consequently, from the tokenizer perspective, the observed performance gains can be attributed to two complementary factors: the transition from hard to soft identifiers and the recommendation-aware refinement of item identifiers.
\end{itemize}

\section{Related Work}
\label{sec:related}

In this section, we review the existing literature relevant to our work, categorizing them into two streams: sequential recommendation and generative recommendation.

\vspace{+3pt}
\textbf{Sequential Recommendation.} Sequential recommendation focuses on predicting a user's next interaction based on their historical behavior sequences. Early methods relied on convolutional neural networks (CNNs)~\cite{Caser} or recurrent neural networks (RNNs)~\cite{GRU4Rec} to capture sequential dependencies. Motivated by advances in natural language processing, Transformer-based models such as SASRec~\cite{SASRec} and BERT4Rec~\cite{BERT4Rec} have become widely adopted, leveraging self-attention to model long-range dependencies efficiently. $\text{S}^3$-Rec~\cite{S3-Rec} further improves performance by maximizing mutual information across multiple types of contextual data. Despite their success, these methods generally follow a pure ID-based paradigm, representing items as randomly initialized atomic identifiers. This design inherently lacks semantic awareness, which restricts knowledge transfer across items, reduces generalization to cold-start scenarios, and limits scalability to large item collections.

\vspace{+3pt}
\textbf{Generative Recommendation.} Generative recommendation has been largely inspired by advances in LLMs~\cite{qwen3, GeneRec,hou2025generative,kuaishou_survey,li-etal-2024-large}. Early methods indexed items using purely text-based identifiers~\cite{P5,IDGenRec,BIGRec}. For example, BIGRec~\cite{BIGRec} represents items using their titles and fine-tunes LLMs for recommendation under a bi-step grounding paradigm. BinLLM~\cite{BinLLM} transforms collaborative embeddings obtained from external models into binary sequences, thereby aligning item representations with formats readily consumable by LLMs. While still compatible with LLM inputs, these approaches struggled to capture item-level collaborative signals~\cite{CoLLM,BinLLM,llara}. Subsequent approaches reformulate recommendation as a sequence-to-sequence task by tokenizing item sequences, providing a framework for jointly modeling semantic relevance and collaborative patterns~\cite{EAGER,SEATER,TIGER,LETTER,DiscRec,LC-Rec}. For example, TIGER~\cite{TIGER} represents each item with a Semantic ID—a tuple of codewords—and uses a Transformer to predict the next one. LETTER~\cite{LETTER} further introduces collaborative and diversity regularization to enhance the quality of the learned identifiers. OneRec~\cite{Onerec-v2} employs RQ-KMeans for tokenization, successfully replacing cascaded pipelines in industrial applications. 

However, most existing methods rely on a rigid two-stage pipeline in which the tokenizer is frozen during recommender training. This disjoint optimization renders the identifiers agnostic to downstream objectives, resulting in suboptimal representations~\cite{BLOGER}. More recent work, such as ETEG-Rec~\cite{ETEGRec}, explores alternating optimization to address this issue but still lacks a unified high-level objective. In contrast, our UniGRec addresses this critical gap in generative recommendation by leveraging soft identifiers, allowing the recommendation loss to jointly supervise both models. Combined with carefully designed training strategies, this enables fully end-to-end alignment between the tokenizer and recommender, bridging the disconnect present in prior two-stage pipelines.
\section{Conclusion}
\label{sec:conclusion}

In this work, we addressed the limitations of disjoint optimization in generative recommendation by proposing UniGRec, a unified framework that jointly integrated the tokenizer and recommender. By leveraging differentiable soft identifiers, UniGRec established a direct gradient pathway, allowing the recommendation objective to supervise the tokenization process end-to-end. To tackle the challenges arising from this design, our framework employed Annealed Inference Alignment to mitigate the training–inference discrepancy, Codeword Uniformity Regularization to prevent identifier collapse, and Dual Collaborative Distillation to reinforce coarse-grained collaborative signals.

For future work, we intend to explore more efficient approaches for integrating collaborative signals in a more streamlined and effective manner, given that our current framework depends on a pre-trained collaborative model and relies on two distillation losses. In addition, we aim to evaluate the scalability of UniGRec on large-scale industrial datasets~\cite{fu2025forge}, in order to further assess its robustness and efficiency when applied to extremely large item corpora.

\bibliographystyle{ACM-Reference-Format}
\balance
\bibliography{8_ref}


\begin{thebibliography}{42}


\ifx \showCODEN    \undefined \def \showCODEN     #1{\unskip}     \fi
\ifx \showISBNx    \undefined \def \showISBNx     #1{\unskip}     \fi
\ifx \showISBNxiii \undefined \def \showISBNxiii  #1{\unskip}     \fi
\ifx \showISSN     \undefined \def \showISSN      #1{\unskip}     \fi
\ifx \showLCCN     \undefined \def \showLCCN      #1{\unskip}     \fi
\ifx \shownote     \undefined \def \shownote      #1{#1}          \fi
\ifx \showarticletitle \undefined \def \showarticletitle #1{#1}   \fi
\ifx \showURL      \undefined \def \showURL       {\relax}        \fi
\providecommand\bibfield[2]{#2}
\providecommand\bibinfo[2]{#2}
\providecommand\natexlab[1]{#1}
\providecommand\showeprint[2][]{arXiv:#2}

\bibitem[Bai et~al\mbox{.}(2025)]%
        {BLOGER}
\bibfield{author}{\bibinfo{person}{Yimeng Bai}, \bibinfo{person}{Chang Liu}, \bibinfo{person}{Yang Zhang}, \bibinfo{person}{Dingxian Wang}, \bibinfo{person}{Frank Yang}, \bibinfo{person}{Andrew Rabinovich}, \bibinfo{person}{Wenge Rong}, {and} \bibinfo{person}{Fuli Feng}.} \bibinfo{year}{2025}\natexlab{}.
\newblock \showarticletitle{Bi-Level Optimization for Generative Recommendation: Bridging Tokenization and Generation}.
\newblock \bibinfo{journal}{\emph{arXiv preprint arXiv:2510.21242}} (\bibinfo{year}{2025}).
\newblock


\bibitem[Bao et~al\mbox{.}(2025)]%
        {BIGRec}
\bibfield{author}{\bibinfo{person}{Keqin Bao}, \bibinfo{person}{Jizhi Zhang}, \bibinfo{person}{Wenjie Wang}, \bibinfo{person}{Yang Zhang}, \bibinfo{person}{Zhengyi Yang}, \bibinfo{person}{Yanchen Luo}, \bibinfo{person}{Chong Chen}, \bibinfo{person}{Fuli Feng}, {and} \bibinfo{person}{Qi Tian}.} \bibinfo{year}{2025}\natexlab{}.
\newblock \showarticletitle{A Bi-Step Grounding Paradigm for Large Language Models in Recommendation Systems}.
\newblock \bibinfo{journal}{\emph{ACM Trans. Recomm. Syst.}} \bibinfo{volume}{3}, \bibinfo{number}{4}, Article \bibinfo{articleno}{53} (\bibinfo{date}{April} \bibinfo{year}{2025}), \bibinfo{numpages}{27}~pages.
\newblock


\bibitem[Baykal et~al\mbox{.}(2024)]%
        {EdVAE}
\bibfield{author}{\bibinfo{person}{Gulcin Baykal}, \bibinfo{person}{Melih Kandemir}, {and} \bibinfo{person}{Gozde Unal}.} \bibinfo{year}{2024}\natexlab{}.
\newblock \showarticletitle{EdVAE: Mitigating codebook collapse with evidential discrete variational autoencoders}.
\newblock \bibinfo{journal}{\emph{Pattern Recognition}}  \bibinfo{volume}{156} (\bibinfo{year}{2024}), \bibinfo{pages}{110792}.
\newblock


\bibitem[Bin et~al\mbox{.}(2025)]%
        {StreamingVQ}
\bibfield{author}{\bibinfo{person}{Xingyan Bin}, \bibinfo{person}{Jianfei Cui}, \bibinfo{person}{Wujie Yan}, \bibinfo{person}{Zhichen Zhao}, \bibinfo{person}{Xintian Han}, \bibinfo{person}{Chongyang Yan}, \bibinfo{person}{Feng Zhang}, \bibinfo{person}{Xun Zhou}, \bibinfo{person}{Xiao Yang}, {and} \bibinfo{person}{Zuotao Liu}.} \bibinfo{year}{2025}\natexlab{}.
\newblock \showarticletitle{Real-time Indexing for Large-scale Recommendation by Streaming Vector Quantization Retriever}. In \bibinfo{booktitle}{\emph{Proceedings of the 31st ACM SIGKDD Conference on Knowledge Discovery and Data Mining V.2}} (Toronto ON, Canada) \emph{(\bibinfo{series}{KDD '25})}. \bibinfo{publisher}{Association for Computing Machinery}, \bibinfo{address}{New York, NY, USA}, \bibinfo{pages}{4273–4283}.
\newblock
\showISBNx{9798400714542}


\bibitem[Cao et~al\mbox{.}(2025)]%
        {OnePiece}
\bibfield{author}{\bibinfo{person}{Jiangxia Cao}, \bibinfo{person}{Shuo Yang}, \bibinfo{person}{Zijun Wang}, {and} \bibinfo{person}{Qinghai Tan}.} \bibinfo{year}{2025}\natexlab{}.
\newblock \showarticletitle{OnePiece: The Great Route to Generative Recommendation--A Case Study from Tencent Algorithm Competition}.
\newblock \bibinfo{journal}{\emph{arXiv preprint arXiv:2512.07424}} (\bibinfo{year}{2025}).
\newblock


\bibitem[Fu et~al\mbox{.}(2025)]%
        {fu2025forge}
\bibfield{author}{\bibinfo{person}{Kairui Fu}, \bibinfo{person}{Tao Zhang}, \bibinfo{person}{Shuwen Xiao}, \bibinfo{person}{Ziyang Wang}, \bibinfo{person}{Xinming Zhang}, \bibinfo{person}{Chenchi Zhang}, \bibinfo{person}{Yuliang Yan}, \bibinfo{person}{Junjun Zheng}, \bibinfo{person}{Yu Li}, \bibinfo{person}{Zhihong Chen}, {et~al\mbox{.}}} \bibinfo{year}{2025}\natexlab{}.
\newblock \showarticletitle{Forge: Forming semantic identifiers for generative retrieval in industrial datasets}.
\newblock \bibinfo{journal}{\emph{arXiv preprint arXiv:2509.20904}} (\bibinfo{year}{2025}).
\newblock


\bibitem[Geng et~al\mbox{.}(2022)]%
        {P5}
\bibfield{author}{\bibinfo{person}{Shijie Geng}, \bibinfo{person}{Shuchang Liu}, \bibinfo{person}{Zuohui Fu}, \bibinfo{person}{Yingqiang Ge}, {and} \bibinfo{person}{Yongfeng Zhang}.} \bibinfo{year}{2022}\natexlab{}.
\newblock \showarticletitle{Recommendation as Language Processing (RLP): A Unified Pretrain, Personalized Prompt \& Predict Paradigm (P5)}. In \bibinfo{booktitle}{\emph{Proceedings of the 16th ACM Conference on Recommender Systems}} (Seattle, WA, USA) \emph{(\bibinfo{series}{RecSys '22})}. \bibinfo{publisher}{Association for Computing Machinery}, \bibinfo{address}{New York, NY, USA}, \bibinfo{pages}{299–315}.
\newblock
\showISBNx{9781450392785}


\bibitem[Han et~al\mbox{.}(2025)]%
        {MTGR}
\bibfield{author}{\bibinfo{person}{Ruidong Han}, \bibinfo{person}{Bin Yin}, \bibinfo{person}{Shangyu Chen}, \bibinfo{person}{He Jiang}, \bibinfo{person}{Fei Jiang}, \bibinfo{person}{Xiang Li}, \bibinfo{person}{Chi Ma}, \bibinfo{person}{Mincong Huang}, \bibinfo{person}{Xiaoguang Li}, \bibinfo{person}{Chunzhen Jing}, \bibinfo{person}{Yueming Han}, \bibinfo{person}{MengLei Zhou}, \bibinfo{person}{Lei Yu}, \bibinfo{person}{Chuan Liu}, {and} \bibinfo{person}{Wei Lin}.} \bibinfo{year}{2025}\natexlab{}.
\newblock \showarticletitle{MTGR: Industrial-Scale Generative Recommendation Framework in Meituan}. In \bibinfo{booktitle}{\emph{Proceedings of the 34th ACM International Conference on Information and Knowledge Management}} (Seoul, Republic of Korea) \emph{(\bibinfo{series}{CIKM '25})}. \bibinfo{publisher}{Association for Computing Machinery}, \bibinfo{address}{New York, NY, USA}, \bibinfo{pages}{5731–5738}.
\newblock
\showISBNx{9798400720406}


\bibitem[Hidasi and Karatzoglou(2018)]%
        {GRU4Rec}
\bibfield{author}{\bibinfo{person}{Bal\'{a}zs Hidasi} {and} \bibinfo{person}{Alexandros Karatzoglou}.} \bibinfo{year}{2018}\natexlab{}.
\newblock \showarticletitle{Recurrent Neural Networks with Top-k Gains for Session-based Recommendations}. In \bibinfo{booktitle}{\emph{Proceedings of the 27th ACM International Conference on Information and Knowledge Management}} (Torino, Italy) \emph{(\bibinfo{series}{CIKM '18})}. \bibinfo{publisher}{Association for Computing Machinery}, \bibinfo{address}{New York, NY, USA}, \bibinfo{pages}{843–852}.
\newblock
\showISBNx{9781450360142}


\bibitem[Hou et~al\mbox{.}(2025a)]%
        {RPG}
\bibfield{author}{\bibinfo{person}{Yupeng Hou}, \bibinfo{person}{Jiacheng Li}, \bibinfo{person}{Ashley Shin}, \bibinfo{person}{Jinsung Jeon}, \bibinfo{person}{Abhishek Santhanam}, \bibinfo{person}{Wei Shao}, \bibinfo{person}{Kaveh Hassani}, \bibinfo{person}{Ning Yao}, {and} \bibinfo{person}{Julian McAuley}.} \bibinfo{year}{2025}\natexlab{a}.
\newblock \showarticletitle{Generating Long Semantic IDs in Parallel for Recommendation}. In \bibinfo{booktitle}{\emph{Proceedings of the 31st ACM SIGKDD Conference on Knowledge Discovery and Data Mining V.2}} (Toronto ON, Canada) \emph{(\bibinfo{series}{KDD '25})}. \bibinfo{publisher}{Association for Computing Machinery}, \bibinfo{address}{New York, NY, USA}, \bibinfo{pages}{956–966}.
\newblock
\showISBNx{9798400714542}


\bibitem[Hou et~al\mbox{.}(2025b)]%
        {hou2025generative}
\bibfield{author}{\bibinfo{person}{Yupeng Hou}, \bibinfo{person}{An Zhang}, \bibinfo{person}{Leheng Sheng}, \bibinfo{person}{Zhengyi Yang}, \bibinfo{person}{Xiang Wang}, \bibinfo{person}{Tat-Seng Chua}, {and} \bibinfo{person}{Julian McAuley}.} \bibinfo{year}{2025}\natexlab{b}.
\newblock \showarticletitle{Generative Recommendation Models: Progress and Directions}. In \bibinfo{booktitle}{\emph{Companion Proceedings of the ACM on Web Conference 2025}}. \bibinfo{pages}{13--16}.
\newblock


\bibitem[Kang and McAuley(2018)]%
        {SASRec}
\bibfield{author}{\bibinfo{person}{Wang-Cheng Kang} {and} \bibinfo{person}{Julian McAuley}.} \bibinfo{year}{2018}\natexlab{}.
\newblock \showarticletitle{Self-attentive sequential recommendation}. In \bibinfo{booktitle}{\emph{2018 IEEE international conference on data mining (ICDM)}}. IEEE, \bibinfo{publisher}{{IEEE} Computer Society}, \bibinfo{pages}{197--206}.
\newblock


\bibitem[Lee et~al\mbox{.}(2022)]%
        {RQ-VAE}
\bibfield{author}{\bibinfo{person}{Doyup Lee}, \bibinfo{person}{Chiheon Kim}, \bibinfo{person}{Saehoon Kim}, \bibinfo{person}{Minsu Cho}, {and} \bibinfo{person}{Wook-Shin Han}.} \bibinfo{year}{2022}\natexlab{}.
\newblock \showarticletitle{Autoregressive Image Generation Using Residual Quantization}. In \bibinfo{booktitle}{\emph{Proceedings of the IEEE/CVF Conference on Computer Vision and Pattern Recognition (CVPR)}}. \bibinfo{pages}{11523--11532}.
\newblock


\bibitem[Li et~al\mbox{.}(2024)]%
        {li-etal-2024-large}
\bibfield{author}{\bibinfo{person}{Lei Li}, \bibinfo{person}{Yongfeng Zhang}, \bibinfo{person}{Dugang Liu}, {and} \bibinfo{person}{Li Chen}.} \bibinfo{year}{2024}\natexlab{}.
\newblock \showarticletitle{Large Language Models for Generative Recommendation: A Survey and Visionary Discussions}. In \bibinfo{booktitle}{\emph{Proceedings of the 2024 Joint International Conference on Computational Linguistics, Language Resources and Evaluation (LREC-COLING 2024)}}. \bibinfo{publisher}{ELRA and ICCL}, \bibinfo{address}{Torino, Italia}, \bibinfo{pages}{10146--10159}.
\newblock


\bibitem[Li et~al\mbox{.}(2025)]%
        {kuaishou_survey}
\bibfield{author}{\bibinfo{person}{Xiaopeng Li}, \bibinfo{person}{Bo Chen}, \bibinfo{person}{Junda She}, \bibinfo{person}{Shiteng Cao}, \bibinfo{person}{You Wang}, \bibinfo{person}{Qinlin Jia}, \bibinfo{person}{Haiying He}, \bibinfo{person}{Zheli Zhou}, \bibinfo{person}{Zhao Liu}, \bibinfo{person}{Ji Liu}, {et~al\mbox{.}}} \bibinfo{year}{2025}\natexlab{}.
\newblock \showarticletitle{A Survey of Generative Recommendation from a Tri-Decoupled Perspective: Tokenization, Architecture, and Optimization}.
\newblock  (\bibinfo{year}{2025}).
\newblock


\bibitem[Liao et~al\mbox{.}(2024)]%
        {llara}
\bibfield{author}{\bibinfo{person}{Jiayi Liao}, \bibinfo{person}{Sihang Li}, \bibinfo{person}{Zhengyi Yang}, \bibinfo{person}{Jiancan Wu}, \bibinfo{person}{Yancheng Yuan}, \bibinfo{person}{Xiang Wang}, {and} \bibinfo{person}{Xiangnan He}.} \bibinfo{year}{2024}\natexlab{}.
\newblock \showarticletitle{LLaRA: Large Language-Recommendation Assistant}. In \bibinfo{booktitle}{\emph{Proceedings of the 47th International ACM SIGIR Conference on Research and Development in Information Retrieval}} (Washington DC, USA) \emph{(\bibinfo{series}{SIGIR '24})}. \bibinfo{publisher}{Association for Computing Machinery}, \bibinfo{address}{New York, NY, USA}, \bibinfo{pages}{1785–1795}.
\newblock
\showISBNx{9798400704314}


\bibitem[Liu et~al\mbox{.}(2025a)]%
        {DiscRec}
\bibfield{author}{\bibinfo{person}{Chang Liu}, \bibinfo{person}{Yimeng Bai}, \bibinfo{person}{Xiaoyan Zhao}, \bibinfo{person}{Yang Zhang}, \bibinfo{person}{Fuli Feng}, {and} \bibinfo{person}{Wenge Rong}.} \bibinfo{year}{2025}\natexlab{a}.
\newblock \showarticletitle{DiscRec: Disentangled Semantic-Collaborative Modeling for Generative Recommendation}.
\newblock \bibinfo{journal}{\emph{arXiv preprint arXiv:2506.15576}} (\bibinfo{year}{2025}).
\newblock


\bibitem[Liu et~al\mbox{.}(2025b)]%
        {ETEGRec}
\bibfield{author}{\bibinfo{person}{Enze Liu}, \bibinfo{person}{Bowen Zheng}, \bibinfo{person}{Cheng Ling}, \bibinfo{person}{Lantao Hu}, \bibinfo{person}{Han Li}, {and} \bibinfo{person}{Wayne~Xin Zhao}.} \bibinfo{year}{2025}\natexlab{b}.
\newblock \showarticletitle{Generative Recommender with End-to-End Learnable Item Tokenization}. In \bibinfo{booktitle}{\emph{Proceedings of the 48th International ACM SIGIR Conference on Research and Development in Information Retrieval}} (Padua, Italy) \emph{(\bibinfo{series}{SIGIR '25})}. \bibinfo{publisher}{Association for Computing Machinery}, \bibinfo{address}{New York, NY, USA}, \bibinfo{pages}{729–739}.
\newblock
\showISBNx{9798400715921}


\bibitem[Loshchilov and Hutter(2019)]%
        {AdamW}
\bibfield{author}{\bibinfo{person}{Ilya Loshchilov} {and} \bibinfo{person}{Frank Hutter}.} \bibinfo{year}{2019}\natexlab{}.
\newblock \showarticletitle{Decoupled Weight Decay Regularization}. In \bibinfo{booktitle}{\emph{7th International Conference on Learning Representations, {ICLR} 2019, New Orleans, LA, USA, May 6-9, 2019}}. \bibinfo{publisher}{OpenReview.net}.
\newblock


\bibitem[Ma et~al\mbox{.}(2019)]%
        {HGN}
\bibfield{author}{\bibinfo{person}{Chen Ma}, \bibinfo{person}{Peng Kang}, {and} \bibinfo{person}{Xue Liu}.} \bibinfo{year}{2019}\natexlab{}.
\newblock \showarticletitle{Hierarchical Gating Networks for Sequential Recommendation}. In \bibinfo{booktitle}{\emph{Proceedings of the 25th ACM SIGKDD International Conference on Knowledge Discovery \& Data Mining}} (Anchorage, AK, USA) \emph{(\bibinfo{series}{KDD '19})}. \bibinfo{publisher}{Association for Computing Machinery}, \bibinfo{address}{New York, NY, USA}, \bibinfo{pages}{825–833}.
\newblock
\showISBNx{9781450362016}


\bibitem[Raffel et~al\mbox{.}(2020)]%
        {T5}
\bibfield{author}{\bibinfo{person}{Colin Raffel}, \bibinfo{person}{Noam Shazeer}, \bibinfo{person}{Adam Roberts}, \bibinfo{person}{Katherine Lee}, \bibinfo{person}{Sharan Narang}, \bibinfo{person}{Michael Matena}, \bibinfo{person}{Yanqi Zhou}, \bibinfo{person}{Wei Li}, {and} \bibinfo{person}{Peter~J Liu}.} \bibinfo{year}{2020}\natexlab{}.
\newblock \showarticletitle{Exploring the limits of transfer learning with a unified text-to-text transformer}.
\newblock \bibinfo{journal}{\emph{Journal of machine learning research}} \bibinfo{volume}{21}, \bibinfo{number}{140} (\bibinfo{year}{2020}), \bibinfo{pages}{1--67}.
\newblock


\bibitem[Rajput et~al\mbox{.}(2023)]%
        {TIGER}
\bibfield{author}{\bibinfo{person}{Shashank Rajput}, \bibinfo{person}{Nikhil Mehta}, \bibinfo{person}{Anima Singh}, \bibinfo{person}{Raghunandan Keshavan}, \bibinfo{person}{Trung Vu}, \bibinfo{person}{Lukasz Heidt}, \bibinfo{person}{Lichan Hong}, \bibinfo{person}{Yi Tay}, \bibinfo{person}{Vinh~Q. Tran}, \bibinfo{person}{Jonah Samost}, \bibinfo{person}{Maciej Kula}, \bibinfo{person}{Ed~H. Chi}, {and} \bibinfo{person}{Maheswaran Sathiamoorthy}.} \bibinfo{year}{2023}\natexlab{}.
\newblock \showarticletitle{Recommender systems with generative retrieval}. In \bibinfo{booktitle}{\emph{Proceedings of the 37th International Conference on Neural Information Processing Systems}} (New Orleans, LA, USA) \emph{(\bibinfo{series}{NIPS '23})}. \bibinfo{publisher}{Curran Associates Inc.}, \bibinfo{address}{Red Hook, NY, USA}, Article \bibinfo{articleno}{452}, \bibinfo{numpages}{17}~pages.
\newblock


\bibitem[Shi et~al\mbox{.}(2025)]%
        {IBQ}
\bibfield{author}{\bibinfo{person}{Fengyuan Shi}, \bibinfo{person}{Zhuoyan Luo}, \bibinfo{person}{Yixiao Ge}, \bibinfo{person}{Yujiu Yang}, \bibinfo{person}{Ying Shan}, {and} \bibinfo{person}{Limin Wang}.} \bibinfo{year}{2025}\natexlab{}.
\newblock \showarticletitle{Scalable Image Tokenization with Index Backpropagation Quantization}. In \bibinfo{booktitle}{\emph{Proceedings of the IEEE/CVF International Conference on Computer Vision (ICCV)}}. \bibinfo{pages}{16037--16046}.
\newblock


\bibitem[Si et~al\mbox{.}(2024)]%
        {SEATER}
\bibfield{author}{\bibinfo{person}{Zihua Si}, \bibinfo{person}{Zhongxiang Sun}, \bibinfo{person}{Jiale Chen}, \bibinfo{person}{Guozhang Chen}, \bibinfo{person}{Xiaoxue Zang}, \bibinfo{person}{Kai Zheng}, \bibinfo{person}{Yang Song}, \bibinfo{person}{Xiao Zhang}, \bibinfo{person}{Jun Xu}, {and} \bibinfo{person}{Kun Gai}.} \bibinfo{year}{2024}\natexlab{}.
\newblock \showarticletitle{Generative Retrieval with Semantic Tree-Structured Identifiers and Contrastive Learning}. In \bibinfo{booktitle}{\emph{Proceedings of the 2024 Annual International ACM SIGIR Conference on Research and Development in Information Retrieval in the Asia Pacific Region}} (Tokyo, Japan) \emph{(\bibinfo{series}{SIGIR-AP 2024})}. \bibinfo{publisher}{Association for Computing Machinery}, \bibinfo{address}{New York, NY, USA}, \bibinfo{pages}{154–163}.
\newblock
\showISBNx{9798400707247}


\bibitem[Sun et~al\mbox{.}(2019)]%
        {BERT4Rec}
\bibfield{author}{\bibinfo{person}{Fei Sun}, \bibinfo{person}{Jun Liu}, \bibinfo{person}{Jian Wu}, \bibinfo{person}{Changhua Pei}, \bibinfo{person}{Xiao Lin}, \bibinfo{person}{Wenwu Ou}, {and} \bibinfo{person}{Peng Jiang}.} \bibinfo{year}{2019}\natexlab{}.
\newblock \showarticletitle{BERT4Rec: Sequential Recommendation with Bidirectional Encoder Representations from Transformer}. In \bibinfo{booktitle}{\emph{Proceedings of the 28th ACM International Conference on Information and Knowledge Management}} (Beijing, China) \emph{(\bibinfo{series}{CIKM '19})}. \bibinfo{publisher}{Association for Computing Machinery}, \bibinfo{address}{New York, NY, USA}, \bibinfo{pages}{1441–1450}.
\newblock
\showISBNx{9781450369763}


\bibitem[Tan et~al\mbox{.}(2024)]%
        {IDGenRec}
\bibfield{author}{\bibinfo{person}{Juntao Tan}, \bibinfo{person}{Shuyuan Xu}, \bibinfo{person}{Wenyue Hua}, \bibinfo{person}{Yingqiang Ge}, \bibinfo{person}{Zelong Li}, {and} \bibinfo{person}{Yongfeng Zhang}.} \bibinfo{year}{2024}\natexlab{}.
\newblock \showarticletitle{IDGenRec: LLM-RecSys Alignment with Textual ID Learning}. In \bibinfo{booktitle}{\emph{Proceedings of the 47th International ACM SIGIR Conference on Research and Development in Information Retrieval}} (Washington DC, USA) \emph{(\bibinfo{series}{SIGIR '24})}. \bibinfo{publisher}{Association for Computing Machinery}, \bibinfo{address}{New York, NY, USA}, \bibinfo{pages}{355–364}.
\newblock
\showISBNx{9798400704314}


\bibitem[Tang and Wang(2018)]%
        {Caser}
\bibfield{author}{\bibinfo{person}{Jiaxi Tang} {and} \bibinfo{person}{Ke Wang}.} \bibinfo{year}{2018}\natexlab{}.
\newblock \showarticletitle{Personalized Top-N Sequential Recommendation via Convolutional Sequence Embedding}. In \bibinfo{booktitle}{\emph{Proceedings of the Eleventh ACM International Conference on Web Search and Data Mining}} (Marina Del Rey, CA, USA) \emph{(\bibinfo{series}{WSDM '18})}. \bibinfo{publisher}{Association for Computing Machinery}, \bibinfo{address}{New York, NY, USA}, \bibinfo{pages}{565–573}.
\newblock
\showISBNx{9781450355810}


\bibitem[Wang et~al\mbox{.}(2024a)]%
        {LETTER}
\bibfield{author}{\bibinfo{person}{Wenjie Wang}, \bibinfo{person}{Honghui Bao}, \bibinfo{person}{Xinyu Lin}, \bibinfo{person}{Jizhi Zhang}, \bibinfo{person}{Yongqi Li}, \bibinfo{person}{Fuli Feng}, \bibinfo{person}{See-Kiong Ng}, {and} \bibinfo{person}{Tat-Seng Chua}.} \bibinfo{year}{2024}\natexlab{a}.
\newblock \showarticletitle{Learnable Item Tokenization for Generative Recommendation}. In \bibinfo{booktitle}{\emph{Proceedings of the 33rd ACM International Conference on Information and Knowledge Management}} (Boise, ID, USA) \emph{(\bibinfo{series}{CIKM '24})}. \bibinfo{publisher}{Association for Computing Machinery}, \bibinfo{address}{New York, NY, USA}, \bibinfo{pages}{2400–2409}.
\newblock
\showISBNx{9798400704369}


\bibitem[Wang et~al\mbox{.}(2025)]%
        {GeneRec}
\bibfield{author}{\bibinfo{person}{Wenjie Wang}, \bibinfo{person}{Xinyu Lin}, \bibinfo{person}{Fuli Feng}, \bibinfo{person}{Xiangnan He}, {and} \bibinfo{person}{Tat-Seng Chua}.} \bibinfo{year}{2025}\natexlab{}.
\newblock \showarticletitle{Generative Recommendation: Towards Personalized Multimodal Content Generation}. In \bibinfo{booktitle}{\emph{Companion Proceedings of the ACM on Web Conference 2025}} (Sydney NSW, Australia) \emph{(\bibinfo{series}{WWW '25})}. \bibinfo{publisher}{Association for Computing Machinery}, \bibinfo{address}{New York, NY, USA}, \bibinfo{pages}{2421–2425}.
\newblock
\showISBNx{9798400713316}


\bibitem[Wang et~al\mbox{.}(2024b)]%
        {EAGER}
\bibfield{author}{\bibinfo{person}{Ye Wang}, \bibinfo{person}{Jiahao Xun}, \bibinfo{person}{Minjie Hong}, \bibinfo{person}{Jieming Zhu}, \bibinfo{person}{Tao Jin}, \bibinfo{person}{Wang Lin}, \bibinfo{person}{Haoyuan Li}, \bibinfo{person}{Linjun Li}, \bibinfo{person}{Yan Xia}, \bibinfo{person}{Zhou Zhao}, {and} \bibinfo{person}{Zhenhua Dong}.} \bibinfo{year}{2024}\natexlab{b}.
\newblock \showarticletitle{EAGER: Two-Stream Generative Recommender with Behavior-Semantic Collaboration}. In \bibinfo{booktitle}{\emph{Proceedings of the 30th ACM SIGKDD Conference on Knowledge Discovery and Data Mining}} (Barcelona, Spain) \emph{(\bibinfo{series}{KDD '24})}. \bibinfo{publisher}{Association for Computing Machinery}, \bibinfo{address}{New York, NY, USA}, \bibinfo{pages}{3245–3254}.
\newblock
\showISBNx{9798400704901}


\bibitem[Xu et~al\mbox{.}(2025)]%
        {MMQ}
\bibfield{author}{\bibinfo{person}{Yi Xu}, \bibinfo{person}{Moyu Zhang}, \bibinfo{person}{Chenxuan Li}, \bibinfo{person}{Zhihao Liao}, \bibinfo{person}{Haibo Xing}, \bibinfo{person}{Hao Deng}, \bibinfo{person}{Jinxin Hu}, \bibinfo{person}{Yu Zhang}, \bibinfo{person}{Xiaoyi Zeng}, {and} \bibinfo{person}{Jing Zhang}.} \bibinfo{year}{2025}\natexlab{}.
\newblock \showarticletitle{MMQ: Multimodal mixture-of-quantization tokenization for semantic id generation and user behavioral adaptation}.
\newblock \bibinfo{journal}{\emph{arXiv preprint arXiv:2508.15281}} (\bibinfo{year}{2025}).
\newblock


\bibitem[Yan et~al\mbox{.}(2024)]%
        {Trinity}
\bibfield{author}{\bibinfo{person}{Jing Yan}, \bibinfo{person}{Liu Jiang}, \bibinfo{person}{Jianfei Cui}, \bibinfo{person}{Zhichen Zhao}, \bibinfo{person}{Xingyan Bin}, \bibinfo{person}{Feng Zhang}, {and} \bibinfo{person}{Zuotao Liu}.} \bibinfo{year}{2024}\natexlab{}.
\newblock \showarticletitle{Trinity: Syncretizing Multi-/Long-Tail/Long-Term Interests All in One}. In \bibinfo{booktitle}{\emph{Proceedings of the 30th ACM SIGKDD Conference on Knowledge Discovery and Data Mining}} (Barcelona, Spain) \emph{(\bibinfo{series}{KDD '24})}. \bibinfo{publisher}{Association for Computing Machinery}, \bibinfo{address}{New York, NY, USA}, \bibinfo{pages}{6095–6104}.
\newblock
\showISBNx{9798400704901}


\bibitem[Yang et~al\mbox{.}(2025)]%
        {qwen3}
\bibfield{author}{\bibinfo{person}{An Yang}, \bibinfo{person}{Anfeng Li}, \bibinfo{person}{Baosong Yang}, \bibinfo{person}{Beichen Zhang}, \bibinfo{person}{Binyuan Hui}, \bibinfo{person}{Bo Zheng}, \bibinfo{person}{Bowen Yu}, \bibinfo{person}{Chang Gao}, \bibinfo{person}{Chengen Huang}, \bibinfo{person}{Chenxu Lv}, {et~al\mbox{.}}} \bibinfo{year}{2025}\natexlab{}.
\newblock \showarticletitle{Qwen3 technical report}.
\newblock \bibinfo{journal}{\emph{arXiv preprint arXiv:2505.09388}} (\bibinfo{year}{2025}).
\newblock


\bibitem[Zhang et~al\mbox{.}(2024)]%
        {BinLLM}
\bibfield{author}{\bibinfo{person}{Yang Zhang}, \bibinfo{person}{Keqin Bao}, \bibinfo{person}{Ming Yan}, \bibinfo{person}{Wenjie Wang}, \bibinfo{person}{Fuli Feng}, {and} \bibinfo{person}{Xiangnan He}.} \bibinfo{year}{2024}\natexlab{}.
\newblock \showarticletitle{Text-like Encoding of Collaborative Information in Large Language Models for Recommendation}. In \bibinfo{booktitle}{\emph{Proceedings of the 62nd Annual Meeting of the Association for Computational Linguistics (Volume 1: Long Papers)}}. \bibinfo{publisher}{Association for Computational Linguistics}, \bibinfo{address}{Bangkok, Thailand}, \bibinfo{pages}{9181--9191}.
\newblock


\bibitem[Zhang et~al\mbox{.}(2025)]%
        {CoLLM}
\bibfield{author}{\bibinfo{person}{Yang Zhang}, \bibinfo{person}{Fuli Feng}, \bibinfo{person}{Jizhi Zhang}, \bibinfo{person}{Keqin Bao}, \bibinfo{person}{Qifan Wang}, {and} \bibinfo{person}{Xiangnan He}.} \bibinfo{year}{2025}\natexlab{}.
\newblock \showarticletitle{CoLLM: Integrating Collaborative Embeddings Into Large Language Models for Recommendation}.
\newblock \bibinfo{journal}{\emph{IEEE Transactions on Knowledge and Data Engineering}} \bibinfo{volume}{37}, \bibinfo{number}{5} (\bibinfo{year}{2025}), \bibinfo{pages}{2329--2340}.
\newblock


\bibitem[Zhao et~al\mbox{.}(2021)]%
        {recbole}
\bibfield{author}{\bibinfo{person}{Wayne~Xin Zhao}, \bibinfo{person}{Shanlei Mu}, \bibinfo{person}{Yupeng Hou}, \bibinfo{person}{Zihan Lin}, \bibinfo{person}{Yushuo Chen}, \bibinfo{person}{Xingyu Pan}, \bibinfo{person}{Kaiyuan Li}, \bibinfo{person}{Yujie Lu}, \bibinfo{person}{Hui Wang}, \bibinfo{person}{Changxin Tian}, \bibinfo{person}{Yingqian Min}, \bibinfo{person}{Zhichao Feng}, \bibinfo{person}{Xinyan Fan}, \bibinfo{person}{Xu Chen}, \bibinfo{person}{Pengfei Wang}, \bibinfo{person}{Wendi Ji}, \bibinfo{person}{Yaliang Li}, \bibinfo{person}{Xiaoling Wang}, {and} \bibinfo{person}{Ji-Rong Wen}.} \bibinfo{year}{2021}\natexlab{}.
\newblock \showarticletitle{RecBole: Towards a Unified, Comprehensive and Efficient Framework for Recommendation Algorithms}. In \bibinfo{booktitle}{\emph{Proceedings of the 30th ACM International Conference on Information \& Knowledge Management}} (Virtual Event, Queensland, Australia) \emph{(\bibinfo{series}{CIKM '21})}. \bibinfo{publisher}{Association for Computing Machinery}, \bibinfo{address}{New York, NY, USA}, \bibinfo{pages}{4653–4664}.
\newblock
\showISBNx{9781450384469}


\bibitem[Zheng et~al\mbox{.}(2024)]%
        {LC-Rec}
\bibfield{author}{\bibinfo{person}{Bowen Zheng}, \bibinfo{person}{Yupeng Hou}, \bibinfo{person}{Hongyu Lu}, \bibinfo{person}{Yu Chen}, \bibinfo{person}{Wayne~Xin Zhao}, \bibinfo{person}{Ming Chen}, {and} \bibinfo{person}{Ji-Rong Wen}.} \bibinfo{year}{2024}\natexlab{}.
\newblock \showarticletitle{Adapting Large Language Models by Integrating Collaborative Semantics for Recommendation}. In \bibinfo{booktitle}{\emph{2024 IEEE 40th International Conference on Data Engineering (ICDE)}}. \bibinfo{pages}{1435--1448}.
\newblock


\bibitem[Zhou et~al\mbox{.}(2025a)]%
        {zhou2025openonerec}
\bibfield{author}{\bibinfo{person}{Guorui Zhou}, \bibinfo{person}{Honghui Bao}, \bibinfo{person}{Jiaming Huang}, \bibinfo{person}{Jiaxin Deng}, \bibinfo{person}{Jinghao Zhang}, \bibinfo{person}{Junda She}, \bibinfo{person}{Kuo Cai}, \bibinfo{person}{Lejian Ren}, \bibinfo{person}{Lu Ren}, \bibinfo{person}{Qiang Luo}, {et~al\mbox{.}}} \bibinfo{year}{2025}\natexlab{a}.
\newblock \showarticletitle{OpenOneRec Technical Report}.
\newblock \bibinfo{journal}{\emph{arXiv preprint arXiv:2512.24762}} (\bibinfo{year}{2025}).
\newblock


\bibitem[Zhou et~al\mbox{.}(2025b)]%
        {Onerec-v2}
\bibfield{author}{\bibinfo{person}{Guorui Zhou}, \bibinfo{person}{Hengrui Hu}, \bibinfo{person}{Hongtao Cheng}, \bibinfo{person}{Huanjie Wang}, \bibinfo{person}{Jiaxin Deng}, \bibinfo{person}{Jinghao Zhang}, \bibinfo{person}{Kuo Cai}, \bibinfo{person}{Lejian Ren}, \bibinfo{person}{Lu Ren}, \bibinfo{person}{Liao Yu}, {et~al\mbox{.}}} \bibinfo{year}{2025}\natexlab{b}.
\newblock \showarticletitle{Onerec-v2 technical report}.
\newblock \bibinfo{journal}{\emph{arXiv preprint arXiv:2508.20900}} (\bibinfo{year}{2025}).
\newblock


\bibitem[Zhou et~al\mbox{.}(2020)]%
        {S3-Rec}
\bibfield{author}{\bibinfo{person}{Kun Zhou}, \bibinfo{person}{Hui Wang}, \bibinfo{person}{Wayne~Xin Zhao}, \bibinfo{person}{Yutao Zhu}, \bibinfo{person}{Sirui Wang}, \bibinfo{person}{Fuzheng Zhang}, \bibinfo{person}{Zhongyuan Wang}, {and} \bibinfo{person}{Ji-Rong Wen}.} \bibinfo{year}{2020}\natexlab{}.
\newblock \showarticletitle{S3-Rec: Self-Supervised Learning for Sequential Recommendation with Mutual Information Maximization}. In \bibinfo{booktitle}{\emph{Proceedings of the 29th ACM International Conference on Information \& Knowledge Management}} (Virtual Event, Ireland) \emph{(\bibinfo{series}{CIKM '20})}. \bibinfo{publisher}{Association for Computing Machinery}, \bibinfo{address}{New York, NY, USA}, \bibinfo{pages}{1893–1902}.
\newblock
\showISBNx{9781450368599}


\bibitem[Zhu et~al\mbox{.}(2025a)]%
        {RankMixer}
\bibfield{author}{\bibinfo{person}{Jie Zhu}, \bibinfo{person}{Zhifang Fan}, \bibinfo{person}{Xiaoxie Zhu}, \bibinfo{person}{Yuchen Jiang}, \bibinfo{person}{Hangyu Wang}, \bibinfo{person}{Xintian Han}, \bibinfo{person}{Haoran Ding}, \bibinfo{person}{Xinmin Wang}, \bibinfo{person}{Wenlin Zhao}, \bibinfo{person}{Zhen Gong}, \bibinfo{person}{Huizhi Yang}, \bibinfo{person}{Zheng Chai}, \bibinfo{person}{Zhe Chen}, \bibinfo{person}{Yuchao Zheng}, \bibinfo{person}{Qiwei Chen}, \bibinfo{person}{Feng Zhang}, \bibinfo{person}{Xun Zhou}, \bibinfo{person}{Peng Xu}, \bibinfo{person}{Xiao Yang}, \bibinfo{person}{Di Wu}, {and} \bibinfo{person}{Zuotao Liu}.} \bibinfo{year}{2025}\natexlab{a}.
\newblock \showarticletitle{RankMixer: Scaling Up Ranking Models in Industrial Recommenders}. In \bibinfo{booktitle}{\emph{Proceedings of the 34th ACM International Conference on Information and Knowledge Management}} (Seoul, Republic of Korea) \emph{(\bibinfo{series}{CIKM '25})}. \bibinfo{publisher}{Association for Computing Machinery}, \bibinfo{address}{New York, NY, USA}, \bibinfo{pages}{6309–6316}.
\newblock
\showISBNx{9798400720406}


\bibitem[Zhu et~al\mbox{.}(2025b)]%
        {Zhu_2025_ICCV}
\bibfield{author}{\bibinfo{person}{Yongxin Zhu}, \bibinfo{person}{Bocheng Li}, \bibinfo{person}{Yifei Xin}, \bibinfo{person}{Zhihua Xia}, {and} \bibinfo{person}{Linli Xu}.} \bibinfo{year}{2025}\natexlab{b}.
\newblock \showarticletitle{Addressing Representation Collapse in Vector Quantized Models with One Linear Layer}. In \bibinfo{booktitle}{\emph{Proceedings of the IEEE/CVF International Conference on Computer Vision (ICCV)}}. \bibinfo{pages}{22968--22977}.
\newblock


\end{thebibliography}

\appendix

\end{document}